\documentclass[preprint,12pt]{elsarticle}
\usepackage{amsmath,amssymb,bm}
\usepackage{booktabs,tabularx,array,multirow}
\usepackage{graphicx}
\usepackage{float}
\usepackage[section]{placeins}
\usepackage{xcolor}
\usepackage{siunitx}
\usepackage{hyperref}
\hypersetup{hidelinks}
\usepackage{microtype}
\usepackage{algorithm}
\usepackage{algpseudocode}

\graphicspath{{figures/}}
\begin{document}
\renewcommand{\figurename}{Fig.}
\begin{frontmatter}

\title{DiffSpecMatch: Differentiable Waveform Inverse Design with Configurable Optimization Modules for Response-Spectrum Matching of Ground Motions}

\author[inst1]{Xiaohu Hu}
\author[inst1]{Su Chen\corref{cor1}}
\author[inst2]{Zhijun Dai}
\author[inst2]{Lei Fu}
\author[inst3]{Zhongxiang Zhang}
\author[inst1,inst2]{Xiaojun Li}
\cortext[cor1]{Corresponding author: Su Chen. E-mail: \href{mailto:chensuchina@126.com}{chensuchina@126.com}.}
\affiliation[inst1]{organization={State Key Laboratory of Bridge Safety and Resilience, Beijing University of Technology},
  city={Beijing}, postcode={100124}, country={China}}
\affiliation[inst2]{organization={Institute of Geophysics, China Earthquake Administration},
  city={Beijing}, postcode={100081}, country={China}}
\affiliation[inst3]{organization={Key Laboratory of Roads and Railway Engineering Safety Control, Ministry of Education, Shijiazhuang Tiedao University},
  city={Shijiazhuang}, postcode={050043}, country={China}}

\begin{abstract}
Response-spectrum matching is an underdetermined waveform inverse problem in which spectral compatibility must be achieved without uncontrolled changes in other engineering-relevant ground-motion characteristics. This study formulates the problem as differentiable waveform inverse design using a fixed SDOF dynamics engine and configurable optimization modules. An envelope-guided parameterization defines the correction waveform, while a hard projection enforces zero correction-induced terminal velocity and displacement. Differentiable SDOF responses then provide gradients for waveform optimization. Spectral fidelity and Arias-intensity accumulation are treated as adaptive objectives. Spectral feasibility, PGA, total Arias intensity, and displacement-path requirements are incorporated as constraints. The differentiable spectrum operator agrees with an exact solution of the equation of motion to within 0.002\%. The complete five-damping configuration achieves a maximum spectral error of 5.00\% over 375 ordinates. On a common benchmark, maximum errors of 2.61\% and 8.47\% are obtained for single- and multi-damping matching, respectively, while retaining the additional waveform constraints. Three-component applications further show that different engineering acceptance requirements can be accommodated through module reconfiguration without changing the forward dynamics.
\end{abstract}

\end{frontmatter}

\section{Introduction}

Response-history analysis requires acceleration records that represent a prescribed seismic demand and remain admissible as dynamical-system inputs. Response-spectrum compatibility is therefore central to the design and qualification of buildings, critical facilities, and their structures, systems, and components \cite{rg160,gb50011,gb50267,gb17741}. A response spectrum, however, retains only the peak responses of a family of linear single-degree-of-freedom oscillators. It neither uniquely determines the underlying waveform nor fixes peak ground acceleration (PGA), the temporal accumulation of Arias intensity, significant duration, or the velocity and displacement obtained by integration. Response-spectrum matching is consequently an underdetermined waveform inverse problem: spectral targets must be approached without allowing uncontrolled changes in other quantities needed for physical interpretation and engineering acceptance. This non-uniqueness also leaves waveform degrees of freedom that can be used to accommodate additional engineering constraints.

Early spectrum-compatible simulation methods adjusted Fourier amplitudes or constructed artificial motions from stochastic representations \cite{tsai1972,levy1976,kaul1978,ahmadi1979,narayana1979,preumont1980,spanos1985}. These approaches established the feasibility of prescribing spectral demand, but frequency-domain amplitude correction provides only indirect control over the timing and localization of waveform changes. Time-domain formulations instead superimpose localized correction functions near response-controlling instants. The multiple-damping work of Lilhanand and Tseng demonstrated simultaneous adjustment to several damping-dependent spectra \cite{lilhanand1987,lilhanand1988}. The RspMatch development by Hancock et al. introduced wavelets compatible with zero final velocity and displacement and extended matching to pseudo-spectral acceleration and multiple damping ratios \cite{hancock2006}. Al Atik and Abrahamson subsequently used Gaussian-tapered cosine adjustments to reduce distortion of the seed motion's nonstationary character \cite{alatik2010}, while Broyden updating improved the solution of the coupled correction equations \cite{broyden2016}. These methods remain influential because they translate spectral discrepancies into effective analytical waveform updates.

Alternative representations have broadened the available ways of distributing correction energy. Wavelet, wavelet-packet, and continuous-wavelet formulations localize modifications in time and frequency \cite{suarez2005,mukherjee2002,giaralis2009,montejo2013,huang2017}, whereas S-transform decomposition retains an explicit physical-frequency coordinate during time--frequency adjustment \cite{mo2022}. Wavelet-based stochastic generation has also been used to reproduce the joint time--frequency structure of fully nonstationary seismic motions \cite{genovese2025}, while response-spectrum-compatible pulse-like motions have been constructed by jointly controlling their amplitude envelope and frequency parameters \cite{chen2022mssp}. Hilbert--Huang and eigenfunction formulations have been used to generate tri-directional, drift-resistant, or mutually weakly correlated components \cite{ni2011,li2016,li2017,yang2019,yang2022}. Optimization-based refinements have minimized worst-case multi-damping discrepancies, selected weights for damping-dependent correction histories, and reduced the number of motions required for dynamic analysis \cite{dai2014,li2023,harati2021}. Dedicated multi-damping studies have further addressed prescribed peak displacement, seed-dependent nonstationarity, and stringent nuclear-engineering compatibility maps \cite{choi2003,zhang2010,zhang2015,cheng2023}. Collectively, this literature shows that no single correction basis is uniformly preferable: the representation must be considered together with the response metric, waveform attributes, and admissibility conditions that define a particular matching task.

The same development has exposed a limitation of treating spectral proximity as the sole objective. Motions with nearly identical response spectra may exhibit materially different energy accumulation, duration, peak amplitudes, time--frequency evolution, or integrated-motion trajectories. Energy-compatible and fully nonstationary simulation has therefore coupled spectral compatibility to Arias-intensity buildup or other temporal and spectral energy descriptors \cite{arias1970,huang2017,wang2021}. More recently, complex-valued wavelet modification has been used to generate bidirectional motions subject to simultaneous spectrum- and energy-compatibility requirements \cite{zhou2025mssp}, reinforcing the need to treat spectral fit and waveform attributes jointly. Duration-sensitive synthesis likewise recognizes that matching spectral ordinates alone does not preserve the temporal extent of strong motion \cite{bommer2004,ji2024}. Baseline-compatible adjustment must further distinguish terminal conditions from intermediate velocity or displacement excursions. Engineering rules add additional heterogeneity: acceptance may be expressed as a symmetric pointwise band, an asymmetric range, a lower envelope, an average-suite condition, a limit on under-target points, a frequency-neighborhood rule, or simultaneous spectra at several damping ratios \cite{rg160,gb50011,gb50267,gb17741}. These requirements cannot be represented faithfully by one mean-square spectral loss.

Existing methods address this heterogeneity in different ways. Many workflows use sequential operations---initial PGA scaling, spectral correction, intensity or duration adjustment, baseline treatment, and post-hoc acceptance screening---whereas selected constrained or hybrid formulations already incorporate parts of the engineering acceptance map, such as spectral lower bounds, under-target counts, duration, and component correlation \cite{dai2014,li2016,li2017,li2023,cheng2023}. This progress is important, but the waveform representation, constraint set, solver, and post-processing logic generally remain task specific. Because repairing one observable can perturb another, increasing the number of periods, damping ratios, and engineering clauses makes the choice of basis functions, grouping rules, relaxation factors, correction order, and stopping criteria increasingly consequential. The unresolved methodological need is therefore not merely another adjustment wave or another multi-damping solver. It is a common computational organization that assigns each requirement an appropriate mathematical role while evaluating every differentiable observable from the same candidate waveform.

Differentiable programming provides the mechanism needed for such an organization. Automatic differentiation can propagate sensitivities through general numerical programs and differentiable physics computations \cite{baydin2018,innes2019,belbuteperes2018}. For response-spectrum matching, this makes it possible to differentiate the oscillator response, spectral extrema, peak acceleration, energy measures, and integrated kinematic paths with respect to waveform parameters. Algebraic invariants can then be imposed by projection, distributed preferences can be expressed as adaptively balanced objectives, continuous thresholds can be enforced as constraints, and discrete or nonsmooth clauses can remain external checks. We therefore couple a fixed differentiable dynamics engine with configurable optimization modules, allowing the engineering acceptance map to change without redesigning the forward solver.

This study develops and evaluates that formulation for single- and multi-damping ground-motion matching. The fixed engine maps unconstrained waveform parameters to an envelope-bounded correction, projects the correction to preserve terminal velocity and displacement increments, and evaluates batched SDOF responses. A configurable library supplies spectral-fidelity and Arias-accumulation objectives, spectral-feasibility and ground-motion-attribute constraints, a pathwise displacement constraint, and external engineering checks. Evaluation includes an independent numerical cross-check of the differentiable spectrum operator, mechanism-focused ablations, an integrated five-damping benchmark, common-data comparisons with the Al Atik procedure on the benchmark introduced by Hancock et al., and two three-component applications with distinct acceptance maps. The contribution is a modular inverse-design architecture in which the forward physics is fixed while objectives, constraints, and checks are selected according to the mathematical form of the task.

\section{Differentiable waveform inverse-design framework}\label{sec:method}

\begin{figure}[H]
\centering
\includegraphics[width=0.98\linewidth]{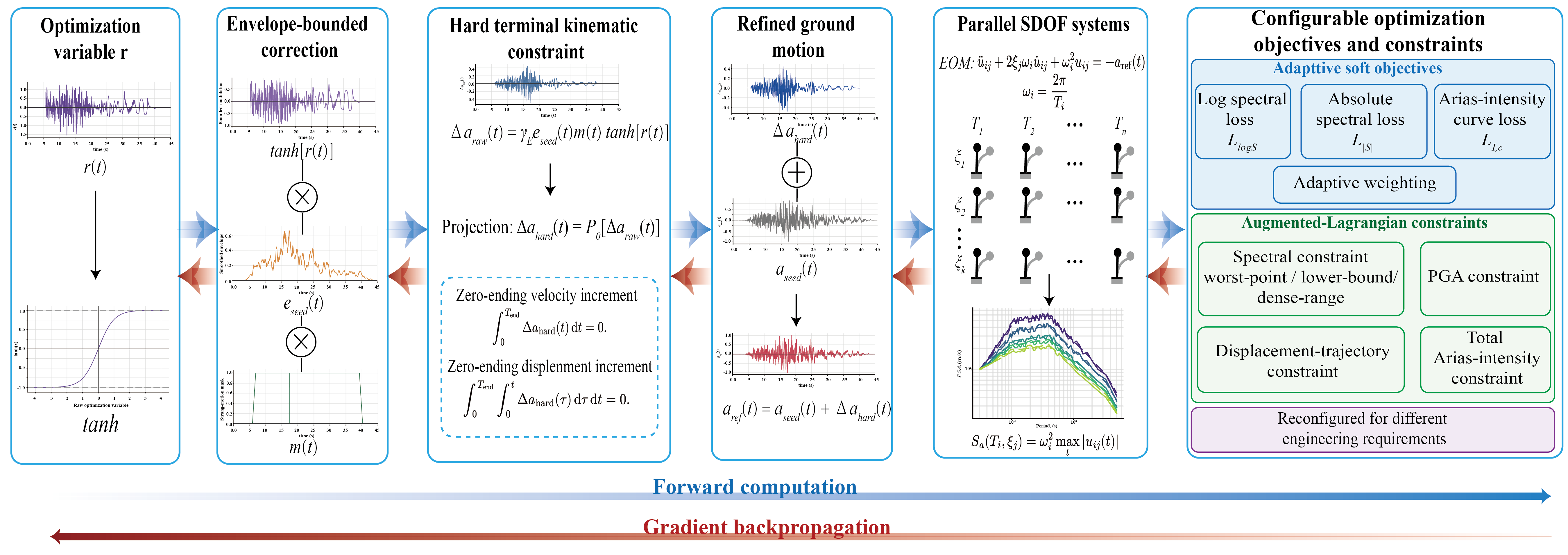}
\caption{Workflow of the differentiable waveform inverse-design method. Waveform parameters are mapped to an envelope-guided bounded correction, projected to enforce zero correction-induced terminal velocity and displacement, and propagated through parallel differentiable SDOF systems. Task-specific objectives and continuous constraints provide gradients, whereas discrete or nonsmooth requirements are evaluated as external acceptance checks.}
\label{fig:workflow}
\end{figure}

Fig.~\ref{fig:workflow} provides an overview of the proposed record-specific inverse-design framework. Given a seed acceleration history and one or more target response spectra, the method seeks waveform parameters whose resulting correction produces an adjusted motion that approaches the spectral targets while satisfying the selected waveform-attribute and kinematic requirements. The framework is organized into a fixed differentiable computational engine and a task-dependent optimization layer. In the fixed engine, the input record is first scaled to the target zero-period ordinate; the waveform parameters are then mapped through a seed-derived envelope and a bounded nonlinear function to define where and by how much the record may be modified. A linear projection subsequently enforces zero correction-induced terminal velocity and displacement, after which the adjusted acceleration is propagated through batched differentiable SDOF systems to obtain the pseudo-spectral accelerations for all prescribed periods and damping ratios. The task-dependent layer assigns different mathematical roles to the quantities evaluated from this common candidate waveform: distributed spectral fidelity and the normalized Arias-intensity accumulation pattern are treated as adaptive objectives; spectral feasibility, PGA, total Arias intensity, and correction-induced displacement are treated as continuous inequality constraints; and discrete or nonsmooth engineering clauses remain external acceptance checks. Adaptive objective balancing and augmented-Lagrangian updates combine the active differentiable modules into a single optimization problem, and the resulting gradients are backpropagated through the SDOF operator, terminal projection, and bounded parameterization to update the waveform parameters directly. Consequently, changing the target damping set or engineering acceptance map does not require redesigning the forward dynamics; it requires only a different selection of active objectives, constraints, and checks.

\subsection{Problem formulation and computational graph}

Each ground-motion record is optimized independently; no model is trained across records. Let $a_0(t_n)$ be the input acceleration sampled at $t_n=n\Delta t$, $n=0,\ldots,N-1$, where $\Delta t$ is the sampling interval, $n$ is the time-sample index, and $N$ is the number of samples. Before optimization, it is scaled to the zero-period target ordinate,
\begin{equation}
  a_s(t)=\frac{a_{\mathrm{PGA}}^{\star}}{\max_t|a_0(t)|}\,a_0(t),
\end{equation}
where $a_{\mathrm{PGA}}^{\star}$ is the target PGA and $a_s(t)$ is the scaled seed acceleration. The adjusted acceleration is
\begin{equation}
  a_r(t)=a_s(t)+\Delta a(t),
\end{equation}
where $a_r(t)$ and $\Delta a(t)$ denote the adjusted acceleration and its correction relative to the scaled seed, respectively. The optimization variable is the waveform-parameter vector $\bm r\in\mathbb R^N$. Bold symbols denote sampled vectors; a tilde denotes a quantity before terminal projection. Gradients are propagated through the computational graph shown in Fig.~\ref{fig:workflow}, where $\bm S_a$ collects the computed spectral ordinates and $L$ is the complete optimization objective. Thus, the correction waveform is determined from the final observables rather than prescribed by a sequence of analytical correction waves.

\subsection{Envelope-guided bounded correction}

An unconstrained correction can place large oscillations in weak-motion intervals. We compute the analytic-signal amplitude $|\mathcal H_a\{a_s\}|$, where $\mathcal H_a\{\cdot\}$ denotes formation of the Hilbert analytic signal, and smooth it with a 0.5-s moving-average window to obtain the seed-derived envelope $E_s(t)$. The active interval is determined from normalized cumulative squared acceleration,
\begin{equation}
 Q_n=\frac{\sum_{k=0}^{n}a_s^2[k]}{\sum_{k=0}^{N-1}a_s^2[k]},
 \qquad 0.002\leq Q_n\leq0.998.
\end{equation}
where $Q_n$ is the fraction of cumulative squared acceleration reached at sample $n$ and $k$ is the summation index. The mask $m_n$ equals one inside this interval and zero outside, with a 100-sample linear taper at both boundaries. The PGA sample and its neighbors remain active. We first construct the pre-projection correction
\begin{equation}\label{eq:param}
 \widetilde{\Delta a}(t)=\gamma_E E_s(t)m(t)\tanh r(t),
\end{equation}
where $r(t)$ is the time-domain representation of $\bm r$, $m(t)$ is the deterministic boundary mask, and $\gamma_E$ is the dimensionless envelope-capacity multiplier. The hyperbolic tangent guarantees
\begin{equation}
 |\widetilde{\Delta a}(t)|\leq \gamma_E E_s(t)m(t).
\end{equation}
The final correction is $\Delta\bm a=\mathcal P_0[\widetilde{\Delta\bm a}]$. The projection preserves the inactive support because its correction basis is multiplied by $\bm m$, but it can slightly alter the pointwise amplitude; the above inequality is therefore a strict bound on the pre-projection correction, not on every sample of the projected correction. The PGA-AL term decides whether a local change around the seed PGA is beneficial.

The envelope multiplier starts at $\gamma_E=1$. Every 500 iterations, the method examines the preceding ten checks. If the spectrum remains infeasible, improvement in normalized spectral violation is below 0.002, and at least one active waveform-correction coefficient satisfies $|\tanh r_n|\geq0.90$, $\gamma_E$ is increased by 0.25, up to 2.5. Thus capacity is enlarged only after the current envelope becomes locally limiting.

\begin{figure}[H]
\centering
\includegraphics[width=0.98\linewidth]{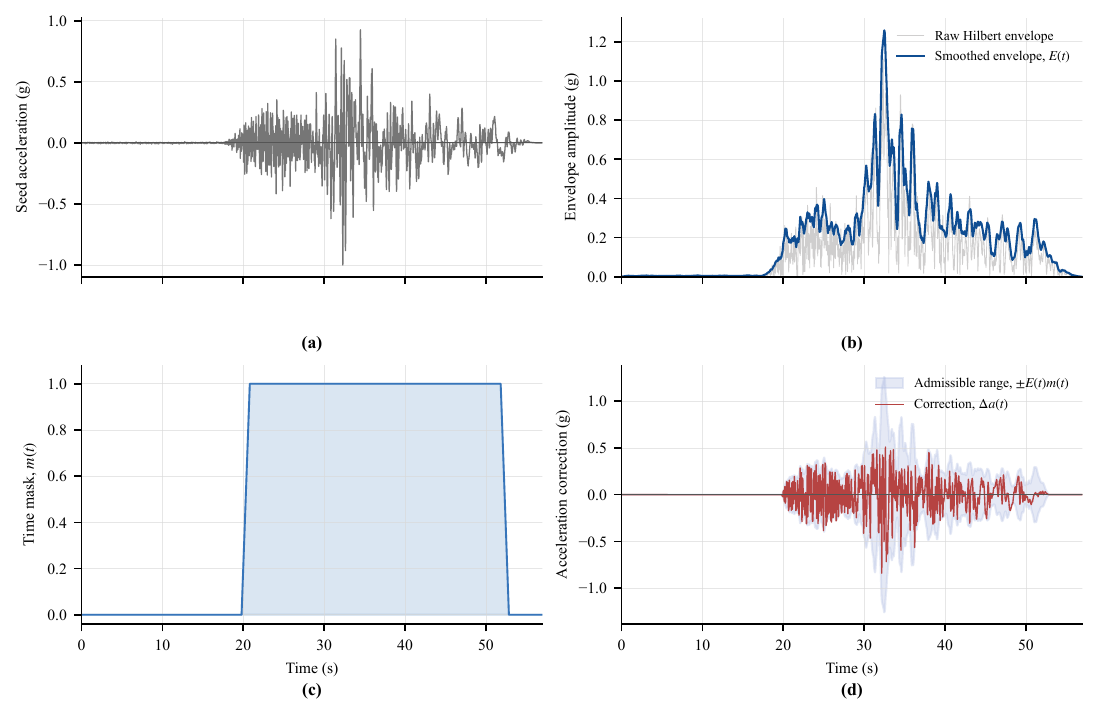}
\caption{From the seed record to an envelope-guided admissible correction. (a) Scaled seed acceleration. (b) Raw Hilbert amplitude and smoothed envelope $E_s(t)$, which sets the local correction scale. (c) Strong-motion mask $m(t)$, which selects the active interval. (d) Envelope-bounded range and the final projected acceleration correction.}
\label{fig:parameterization}
\end{figure}

The four panels in Fig.~\ref{fig:parameterization} connect the seed record to the admissible correction in the order used by the parameterization. The scaled acceleration first identifies the temporal structure of the motion [Fig.~\ref{fig:parameterization}(a)]. Its Hilbert amplitude is then smoothed to obtain the seed-dependent envelope $E_s(t)$ [Fig.~\ref{fig:parameterization}(b)]. Rather than prescribing the correction waveform, this envelope supplies a time-varying amplitude scale. It therefore permits greater adjustment during strong motion and suppresses unnecessary modification during weak-motion intervals.

The principal active interval is identified from the cumulative squared acceleration and represented by the time mask $m(t)$ in Fig.~\ref{fig:parameterization}(c). The mask restricts optimization to the main energy-release portion of the record while retaining the neighborhood of the PGA sample, so that peak acceleration can still be adjusted when required. The unconstrained waveform parameters are subsequently mapped through $\tanh(\cdot)$ and modulated by $E_s(t)m(t)$, producing the envelope-guided correction shown in Fig.~\ref{fig:parameterization}(d). The resulting modification is concentrated mainly within the strong-motion interval, and its admissible magnitude follows the local amplitude of the seed motion. Corrections outside the active region are effectively suppressed. The parameterization therefore does not prescribe a particular correction waveform. It defines where the record may be modified and the approximate magnitude of that modification, while gradients from the active spectral, PGA, Arias-intensity, and kinematic modules determine the detailed waveform adjustment.

The envelope in Fig.~\ref{fig:parameterization}(d) is the strict amplitude bound for the pre-projection correction $\widetilde{\Delta a}(t)$. Because the subsequent hard terminal velocity--displacement projection can slightly modify individual samples, the projected correction $\Delta a(t)$ is not required to satisfy this bound exactly at every time point.

\subsection{Hard terminal velocity--displacement projection}

Double integration makes even a small low-frequency bias consequential. Let $t_f=(N-1)\Delta t$ denote the record duration, and let $\Delta v$ and $\Delta d$ denote the velocity and displacement changes induced solely by $\Delta a$. To prevent the correction from changing the seed record's terminal velocity and displacement, the correction must satisfy
\begin{equation}\label{eq:moments}
 \Delta v(t_f)=\int_0^{t_f}\Delta a(t)\,\mathrm dt=0,
 \qquad
 \Delta d(t_f)=\int_0^{t_f}(t_f-t)\Delta a(t)\,\mathrm dt=0.
\end{equation}
The implementation uses trapezoidal integration. Let $\bm c_v$ and $\bm c_d$ contain the corresponding discrete terminal-velocity and terminal-displacement weights, and define
\begin{equation}
 \bm C=\begin{bmatrix}\bm c_v^{\mathsf T}\\\bm c_d^{\mathsf T}\end{bmatrix},
 \qquad
  \bm B=\operatorname{diag}(\bm m)\bm C^{\mathsf T},
\end{equation}
where $\bm C$ collects the two terminal-increment functionals, $\operatorname{diag}(\bm m)$ is the diagonal matrix formed from the sampled mask, and $\bm B$ restricts the projection basis to its active support. Every candidate is projected as
\begin{equation}\label{eq:projection}
 \Delta\bm a
 =\mathcal P_0[\widetilde{\Delta\bm a}]
 =\widetilde{\Delta\bm a}
 -\bm B(\bm C\bm B)^{-1}\bm C\widetilde{\Delta\bm a}.
\end{equation}
Here, $\mathcal P_0$ denotes the terminal-increment projection operator. Consequently, $\bm C\Delta\bm a=\bm0$ to numerical precision while the projection adjustment remains inside the active mask. This projection enforces $v_r(t_f)=v_s(t_f)$ and $d_r(t_f)=d_s(t_f)$, where $v_s,d_s$ and $v_r,d_r$ are the velocity and displacement histories of the scaled seed and adjusted motions, respectively. Equivalently, the correction contributes zero terminal velocity and displacement increments. It does not guarantee a small displacement deviation at intermediate times and is not a baseline-correction operator for the seed motion. Near-zero adjusted terminal values therefore require a seed record that has already been appropriately baseline corrected.

\subsection{Differentiable SDOF response-spectrum operator}\label{sec:spectrumoperator}

For oscillator period $T_i$ and damping ratio $\zeta_j$, the relative displacement $u_{ij}$ satisfies
\begin{equation}
 \ddot u_{ij}+2\zeta_j\omega_i\dot u_{ij}+\omega_i^2u_{ij}=-a_r(t),
 \qquad \omega_i=\frac{2\pi}{T_i}.
\end{equation}
where $i$ and $j$ index the nonzero oscillator periods and damping ratios, respectively, and $\omega_i$ is the corresponding circular frequency. The pseudo-spectral acceleration is $S_a(T_i,\zeta_j)=\omega_i^2\max_t|u_{ij}(t)|$, with zero initial relative displacement and velocity. For each $(T_i,\zeta_j)$, the discrete impulse response associated with piecewise-linear ground acceleration is precomputed. The displacement history is evaluated as the causal convolution
\begin{equation}
 u_{ij}[n]=\sum_{k=0}^{n}h_{ij}[n-k]a_r[k]+u_{ij}^{(0)}[n],
 \label{eq:fft_convolution}
\end{equation}
where $h_{ij}$ is the discrete oscillator kernel, $k$ is the convolution index, and $u_{ij}^{(0)}$ accounts for the first acceleration sample. Zero-padded fast Fourier transforms evaluate the linear convolution in batches over periods and damping ratios. The recurrence is exact for the adopted time step and within-step interpolation, but the complete spectrum remains a discrete numerical quantity. The maximum operation is piecewise differentiable, and automatic differentiation propagates a subgradient through the controlling response sample. The zero-period ordinate is excluded from the oscillator loss and treated through the PGA module.

The relative error at a nonzero control period is
\begin{equation}
 e_{ij}=\frac{S_a(T_i,\zeta_j)}{S_a^{\star}(T_i,\zeta_j)}-1.
\end{equation}
where $S_a^{\star}(T_i,\zeta_j)$ is the prescribed target ordinate and $e_{ij}$ is the signed relative spectral error. The operator is implemented independently of the optimization layer, which permits a direct numerical cross-check in Section~\ref{sec:operatorvalidation}.

\subsection{Configurable objective and constraint library}

The library separates modules by the observable they control rather than by the optimizer used to enforce them. Distributed preferences are represented as smooth objectives, continuous admissibility thresholds as inequality constraints, and algebraic terminal conditions by the projection in Section~\ref{sec:method}. Only the modules required by a given task are activated.

\subsubsection{Spectral-fidelity objectives}

Two complementary spectral objectives provide distributed optimization signals:
\begin{align}
 L_{\log S}&=\frac{1}{N_S}\sum_{i,j}\left[\log\frac{S_a(T_i,\zeta_j)}{S_a^{\star}(T_i,\zeta_j)}\right]^2,\\
 L_{|S|}&=\frac{1}{N_S}\sum_{i,j}\left[S_a(T_i,\zeta_j)-S_a^{\star}(T_i,\zeta_j)\right]^2.
\end{align}
Here, $N_S$ is the number of nonzero period--damping ordinates. $L_{\log S}$ distributes attention across ordinates of different magnitude, whereas $L_{|S|}$ retains sensitivity to large absolute discrepancies. All period--damping ordinates are equally weighted in both means. Although $L_{|S|}$ is dimensional, its normalized contribution to the combined objective is dimensionless.

\subsubsection{Arias-accumulation objective}

The temporal pattern of energy accumulation is represented by
\begin{equation}
 L_{I,c}=\frac{1}{N}\sum_n\left[C_r(t_n)-C_s(t_n)\right]^2,
\end{equation}
where subscripts $s$ and $r$ identify the scaled seed and adjusted motions, respectively, and $C(t)$ is the normalized cumulative Arias-intensity curve,
\begin{equation}
 I_A(t)=\frac{\pi}{2g}\int_0^t a^2(\tau)\,\mathrm d\tau,
 \qquad C(t)=\frac{I_A(t)}{I_A(t_f)}.
\end{equation}
where $g$ is gravitational acceleration, $\tau$ is the integration variable, $I_A(t)$ is cumulative Arias intensity, and $I_A(t_f)$ is its total value over the record.
Because $C(t)$ is normalized by its final value, matching $C_r(t)$ to $C_s(t)$ controls the relative timing of Arias-intensity accumulation but not its final value:
\begin{equation}
 C_r(t)\simeq C_s(t)\ \nRightarrow\ I_{A,r}(t_f)\simeq I_{A,s}(t_f).
\end{equation}
The separate total-Arias constraint introduced below is therefore necessary when the final Arias intensity must also remain within tolerance.

\subsubsection{Spectral-feasibility constraints}

The signed error $e_{ij}$ supports several acceptance geometries without changing the SDOF operator. Symmetric pointwise closeness, a lower spectral bound, and an asymmetric interval are written, respectively, as
\begin{equation}
 |e_{ij}|\leq\epsilon_S,\qquad e_{ij}\geq-\epsilon_L,\qquad
 -\epsilon_L\leq e_{ij}\leq\epsilon_U,
\end{equation}
where $\epsilon_S$, $\epsilon_L$, and $\epsilon_U$ are the corresponding relative tolerances. For symmetric matching, the continuously differentiated violation jointly covers the $K$ current worst ordinates,
\begin{equation}\label{eq:spectral_violation}
 g_S=\left(\frac{1}{K}\sum_{(i,j)\in\mathcal W_K}
  \left[\frac{|e_{ij}|}{\epsilon_S}-1\right]_+^2\right)^{1/2},
 \qquad [x]_+=\max(x,0),
\end{equation}
where $\mathcal W_K$ is the index set of the $K$ largest absolute errors and $[x]_+$ is the positive-part operator. Analogous one-sided maps are used for lower-bound and asymmetric-range requirements. Joint control of several extreme ordinates discourages the optimizer from repeatedly exchanging one isolated worst point for another.

\subsubsection{Ground-motion-attribute constraints}

PGA and total Arias intensity are controlled independently of their distributed counterparts through
\begin{align}
 g_{\mathrm{PGA}}&=\left[\frac{\left|\max_t|a_r(t)|/a_{\mathrm{PGA}}^{\star}-1\right|}{\epsilon_{\mathrm{PGA}}}-1\right]_+,\\
 g_I&=\left[\frac{|I_A(a_r)/I_A(a_s)-1|}{\epsilon_I}-1\right]_+,
\end{align}
where $\epsilon_{\mathrm{PGA}}$ and $\epsilon_I$ are prescribed relative tolerances and $I_A(a)$ denotes the final Arias intensity of acceleration history $a$. The PGA constraint acts on the zero-period ordinate, which is excluded from the oscillator loss; the total-Arias constraint controls magnitude, whereas $L_{I,c}$ controls normalized accumulation shape.

\subsubsection{Displacement-path constraint}

Zero terminal increments do not preclude a large intermediate excursion followed by a compensating return. The correction-induced displacement path is
\begin{equation}
 \Delta d(t)=\int_0^t\int_0^\tau \Delta a(s)\,\mathrm ds\,\mathrm d\tau,
\end{equation}
with the scale
\begin{equation}
 s_d=\max\left\{\max_t|d_s(t)|,
 \max_{i,j}\frac{S_a^{\star}(T_i,\zeta_j)}{\omega_i^2},10^{-8}\right\}.
\end{equation}
In this definition, $s_d$ is a reference displacement scale formed from the larger of the seed peak displacement, the largest target pseudo-spectral displacement, and a small numerical floor. The pathwise admissibility condition is
\begin{equation}\label{eq:disp_constraint}
 \frac{\max_t|\Delta d(t)|}{\eta_d^{(k)} s_d}-1\leq0.
\end{equation}
Here, $\eta_d^{(k)}$ is the current dimensionless admissibility coefficient for the correction-induced displacement relative to $s_d$. It begins from a restrictive value and is increased only when the displacement constraint remains infeasible over a prescribed patience interval, subject to an upper bound. The resulting final coefficient can therefore differ among seed records. This continuous constraint complements the terminal projection because it acts over the complete displacement path rather than only at $t_f$.

\subsubsection{Engineering-rule configuration}

Engineering requirements are partitioned as
\begin{equation}
 \mathcal R=\mathcal R_{\mathrm{diff}}\cup\mathcal R_{\mathrm{check}},
\end{equation}
where $\mathcal R_{\mathrm{diff}}$ contains requirements represented by continuous or subdifferentiable violation measures with informative waveform gradients, and $\mathcal R_{\mathrm{check}}$ contains suite-level or diagnostic rules evaluated after record-wise optimization. Some code-oriented spectral clauses, such as an under-target count or a frequency-neighborhood condition, use a differentiable surrogate during optimization and their exact rule at periodic feasibility checkpoints. They therefore belong to $\mathcal R_{\mathrm{diff}}$ in the present implementation rather than to post-generation external verification. The response-spectrum operator, waveform parameterization, and terminal projection remain unchanged when the active requirement set changes. Table~\ref{tab:modulelibrary} summarizes the mapping; numerical thresholds and code clauses are stated with the corresponding applications in Section~\ref{sec:applications}.

\begin{table}[H]
\caption{Mapping of response-spectrum matching requirements to optimization and acceptance modules.}
\label{tab:modulelibrary}
\centering\small
\resizebox{\linewidth}{!}{%
\begin{tabular}{p{3.0cm}p{3.1cm}p{3.6cm}p{3.3cm}}
\toprule
Requirement class & Mathematical role & Representative quantity & Treatment\\
\midrule
Distributed spectral fidelity & Smooth preference & $L_{\log S}$, $L_{|S|}$ & Adaptive objective\\
Arias accumulation pattern & Smooth preference & $L_{I,c}$ & Adaptive objective\\
Spectral feasibility & Continuous or subdifferentiable threshold & Symmetric band, lower bound, asymmetric range, count/sum/neighborhood surrogate & Augmented Lagrangian with exact checkpoint test\\
Ground-motion attributes & Continuous threshold & PGA, total Arias intensity & Augmented Lagrangian\\
Terminal compatibility & Linear equality & $\Delta v(t_f)=\Delta d(t_f)=0$ & Hard projection\\
Pathwise kinematics & Continuous threshold & $\max_t|\Delta d(t)|$ & Augmented Lagrangian\\
Suite-level engineering clauses & Post-generation acceptance rule & Significant duration, inter-component correlation & External check\\
\bottomrule
\end{tabular}
}
\end{table}

\subsection{Constrained optimization}

\subsubsection{Adaptive objective balancing}

Fixed weights are poorly transferable because loss magnitudes and gradient scales change with the seed, target, and optimization stage. For $q\in\{\log S,|S|,I_c\}$, let $q$ index the active soft objectives and let superscript $(k)$ denote the optimization iteration. The exponentially smoothed loss scale is
\begin{equation}
 \mu_q^{(k)}=\beta\mu_q^{(k-1)}+(1-\beta)
 \max\{L_q^{(k)},L_q^{\mathrm{ref}}\},\qquad\beta=0.99,
\end{equation}
where $\mu_q^{(k)}$ is the moving loss scale, $\beta$ is its smoothing factor, and $L_q^{\mathrm{ref}}$ is a positive reference floor, with $\mu_q^{(0)}=\max\{L_q^{(0)},L_q^{\mathrm{ref}}\}$. The reference floors are $10^{-12}$ for both spectrum losses and $10^{-3}$ for $L_{I,c}$. The Arias-curve floor is required because the scaled seed has zero normalized Arias-curve mismatch at initialization. The normalized loss is $\widetilde L_q=L_q/\mu_q$. Every 100 iterations, define
\begin{align}
 g_q&=\left\|\nabla_{\bm r}(w_q\widetilde L_q)\right\|_2,\\
 R_q&=\frac{\max\{L_q,L_q^{\mathrm{ref}}\}}
 {\max\{L_q^{(0)},L_q^{\mathrm{ref}}\}},\qquad
 p_q=\frac{R_q}{Q^{-1}\sum_p R_p},
\end{align}
where $w_q$ is the current objective weight, $g_q$ is its gradient norm with respect to $\bm r$, $R_q$ is its loss relative to initialization, $p_q$ is the normalized relative-progress factor, and $Q$ is the number of active soft objectives; $p$ is a dummy objective index and $\bar g=Q^{-1}\sum_qg_q$ is the mean gradient norm. The use of the same reference floor in both the numerator and denominator keeps the relative-progress measure finite even when an objective is initially zero. The preference update is
\begin{equation}
 w_q\leftarrow w_q\left(\frac{\bar g\,p_q^{\alpha}}{g_q}\right)^{\eta_w},
 \quad \alpha=0.5,\quad \eta_w=0.1,
\end{equation}
where $\alpha$ controls the emphasis on relative progress and $\eta_w$ is the weight-update relaxation exponent. The weights are then clipped to $[0.5,2.0]$ and normalized so that $\sum_qw_q=Q$. The soft objective is
\begin{equation}
 L_{\mathrm{soft}}=\frac{1}{Q}\sum_qw_q\widetilde L_q.
\end{equation}

\subsubsection{Augmented-Lagrangian enforcement of continuous constraints}

Soft mean losses do not reliably control an extreme ordinate or a thresholded waveform attribute. For each active continuous requirement indexed by $\ell$, we therefore define a non-negative normalized violation $g_\ell(\bm r)\geq0$, with $g_\ell=0$ denoting feasibility, and add
\begin{equation}\label{eq:al}
 L_{\mathrm{AL},\ell}=\lambda_\ell g_\ell+\frac{\rho_\ell}{2}g_\ell^2.
\end{equation}
Here, $\lambda_\ell$ and $\rho_\ell$ are the Lagrange multiplier and quadratic-penalty coefficient for constraint $\ell$. They are initialized as $\lambda_\ell=0$ and $\rho_\ell=0.1$ and updated every 500 iterations:
\begin{equation}
 \lambda_\ell\leftarrow\min\{10^4,\max(0,\lambda_\ell+\rho_\ell g_\ell)\}.
\end{equation}
If $g_\ell^{(m)}>0.75g_\ell^{(m-1)}$, where superscript $(m)$ indexes AL updates, the penalty is increased as $\rho_\ell\leftarrow\min(1.25\rho_\ell,100)$. The complete objective is
\begin{equation}
 L=L_{\mathrm{soft}}+\sum_\ell L_{\mathrm{AL},\ell}.
\end{equation}

The active $g_\ell$ are selected from the module library: spectral feasibility in Eq.~\eqref{eq:spectral_violation}, PGA and total Arias intensity, and the displacement-path condition in Eq.~\eqref{eq:disp_constraint}. These inequalities express selected admissibility requirements rather than physical conservation laws. Only constraints relevant to the corresponding acceptance protocol enter the augmented Lagrangian.

\subsubsection{Gradient update and convergence}

The waveform parameters are initialized as $\bm r=\bm0$ and updated with Adam using gradients through the complete computational graph. Optimization terminates after sustained satisfaction of the active differentiable constraints and, for code-oriented spectral maps, their associated checkpoint-level feasibility tests. Suite-level external checks are subsequently applied to the generated component waveforms and do not participate in record-wise stopping. If optimization feasibility is not attained within the prescribed computational budget, the best stored checkpoint is reported as non-converged. Algorithm~\ref{alg:inverse-design} summarizes how the parameterization, differentiable dynamics, adaptive objectives, continuous constraints, and checkpoint-level acceptance rules are assembled into the complete optimization procedure.

\begin{algorithm}[H]
\caption{Configurable differentiable waveform inverse-design procedure.}
\label{alg:inverse-design}
\small
\begin{algorithmic}[1]
\Require Seed acceleration $a_0(t)$; target spectra $S_a^\star(T,\zeta)$; target PGA $a_{\mathrm{PGA}}^\star$; active objective set $\mathcal O$; active constraint set $\mathcal C$; engineering acceptance rules $\mathcal R$
\Ensure Adjusted acceleration $a_r(t)$ and its response-spectrum and waveform-attribute evaluations
\State Scale $a_0(t)$ to $a_{\mathrm{PGA}}^\star$ using Eq.~(1) to obtain $a_s(t)$
\State Construct the seed-derived envelope $E_s(t)$ and active mask $m(t)$
\State Initialize $\bm r\gets\bm0$, $\gamma_E\gets1$, $\eta_d\gets0.30$, $w_q\gets1$, $\lambda_\ell\gets0$, $\rho_\ell\gets0.1$, and feasible-check counter $c\gets0$
\For{$k=1,\ldots,K_{\max}$}
  \State Compute $\widetilde{\Delta a}(t)=\gamma_EE_s(t)m(t)\tanh r(t)$
  \State Apply the hard terminal projection $\Delta\bm a\gets\mathcal P_0[\widetilde{\Delta\bm a}]$
  \State Form $\bm a_r\gets\bm a_s+\Delta\bm a$ and compute $S_a(T,\zeta)$ with the differentiable SDOF operator
  \State Evaluate the active soft objectives $L_q$, update their moving scales $\mu_q$, and evaluate the active continuous violations $g_\ell$
  \State Form $L=L_{\mathrm{soft}}+\sum_\ell L_{\mathrm{AL},\ell}$
  \State Backpropagate $\nabla_{\bm r}L$ and update $\bm r$ using Adam with gradient clipping
  \If{$k\bmod100=0$}
    \State Update $w_q$ using Eqs.~(26)--(28) and evaluate the exact checkpoint-level acceptance rules
    \State $c\gets c+1$ if all active requirements are satisfied; otherwise $c\gets0$
    \State Store the current feasible or best checkpoint
  \EndIf
  \If{$k\bmod500=0$}
    \State Update $\lambda_\ell$ and $\rho_\ell$ using Eq.~(31)
    \State If envelope capacity is limiting, set $\gamma_E\gets\min(\gamma_E+0.25,2.5)$
  \EndIf
  \If{the displacement-path constraint remains infeasible after its patience interval}
    \State $\eta_d\gets\min(\eta_d+0.10,1.50)$
  \EndIf
  \If{$c\geq5$}
    \State \textbf{break}
  \EndIf
\EndFor
\State Select the best feasible checkpoint; if none is feasible, minimize active violation first and the soft objective second
\end{algorithmic}
\end{algorithm}

Adam used a learning rate of $5\times10^{-4}$ for the five-damping mechanism study and $10^{-3}$ for the code-oriented applications. The parameter-gradient norm was clipped at 5, and all calculations were performed in single precision with a maximum budget of 100,000 iterations. Adaptive objective weights and checkpoint-level acceptance were evaluated every 100 iterations, whereas the augmented-Lagrangian variables and envelope capacity were updated every 500 iterations as specified above.

\section{Numerical verification and mechanism studies}\label{sec:mechanism}

The numerical assessment begins with an optimization-independent check of the differentiable response-spectrum operator. Controlled ablations then determine whether the Arias, extreme-ordinate, PGA, terminal, and pathwise modules affect their intended observables, after which the complete configuration is evaluated on a dense five-damping target. The mechanism studies use the RG~1.60 horizontal spectral family represented by the control coefficients in GB~50267 Appendix~F~\cite{rg160,gb50267}. Its zero-period ordinate is $1.0g$, and five damping ratios (0.5\%, 2\%, 5\%, 7\%, and 10\%) are prescribed at 75 nonzero periods from 0.0303 to 5.0~s.

\subsection{Numerical verification of the differentiable response-spectrum operator}\label{sec:operatorvalidation}

\begin{figure}[H]
\centering
\includegraphics[width=\linewidth]{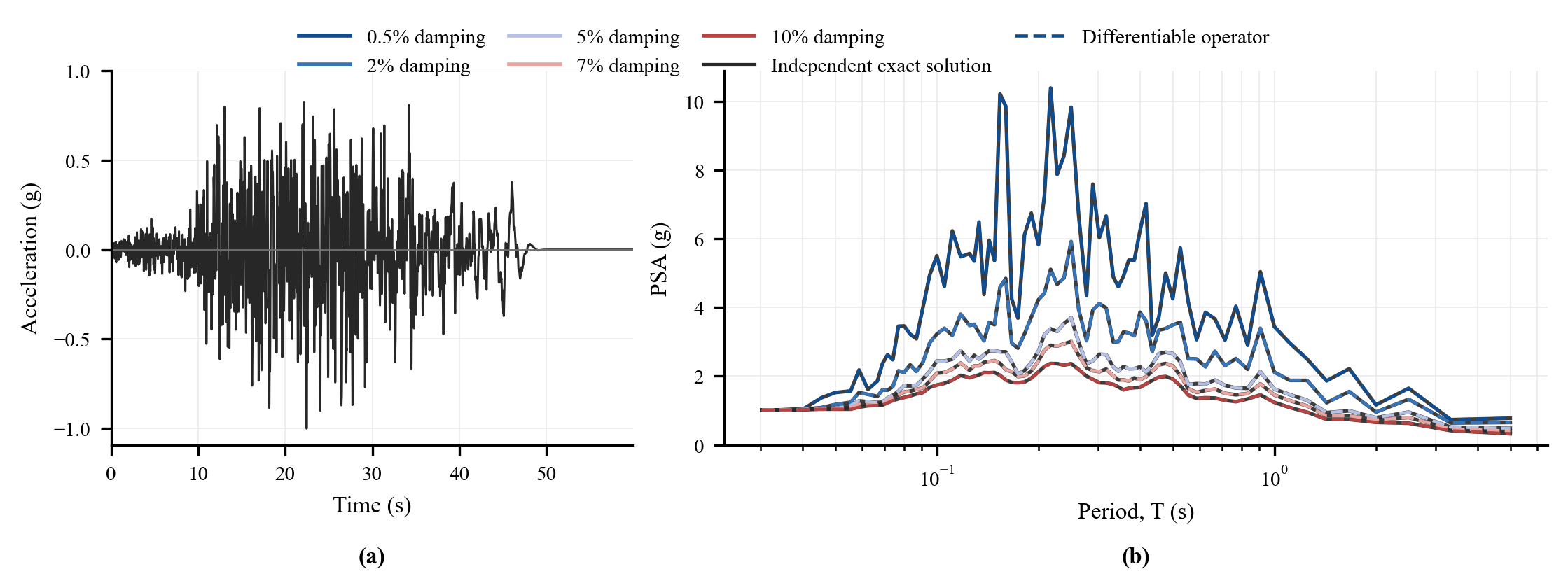}
\caption{Forward-accuracy check for the differentiable response-spectrum operator. (a) Input ground-motion acceleration history. (b) Pseudo-spectral accelerations computed by the differentiable operator and an independent state-transition solution at five damping ratios. Their near-complete overlap, with the black reference curves beneath the colored dashed curves, establishes the numerical agreement used in subsequent optimization.}
\label{fig:spectrumcrosscheck}
\end{figure}

Because every optimization gradient passes through the response-spectrum operator, its forward accuracy must be established before the optimization results are interpreted. Fig.~\ref{fig:spectrumcrosscheck}(a) shows the randomly selected NGA-West2 record used for this independent check. For this input, pseudo-spectral accelerations at damping ratios of 0.5\%, 2\%, 5\%, 7\%, and 10\% were computed using both the differentiable operator and an independent exact solution of the equation of motion~\cite{chopra2001}. The two sets of curves are visually indistinguishable across the full period range [Fig.~\ref{fig:spectrumcrosscheck}(b)]. More quantitatively, the maximum relative difference among all 375 period--damping ordinates is 0.002\%. This agreement validates the forward calculation used in the subsequent gradient-based inverse design.

\subsection{Configurable-module ablation}

Three configurations are compared: the complete method, removal of both Arias terms, and joint removal of the worst-point spectrum and PGA constraints.

For $N_S$ nonzero spectral ordinates, the mean absolute relative error (MARE), root-mean-square relative error (RMS), and maximum absolute relative error are
\begin{align}
 \mathrm{MARE}&=\frac{100}{N_S}\sum_{i,j}|e_{ij}|,\\
 \mathrm{RMS}&=100\sqrt{\frac{1}{N_S}\sum_{i,j}e_{ij}^2},\\
 E_{\max}&=100\max_{i,j}|e_{ij}|.
\end{align}
PGA and total Arias-intensity errors are reported as absolute relative differences. These metrics are used below to distinguish distributed spectral fit from the worst ordinate and from non-spectral attributes.

\begin{figure}[H]
\centering
\includegraphics[width=\linewidth]{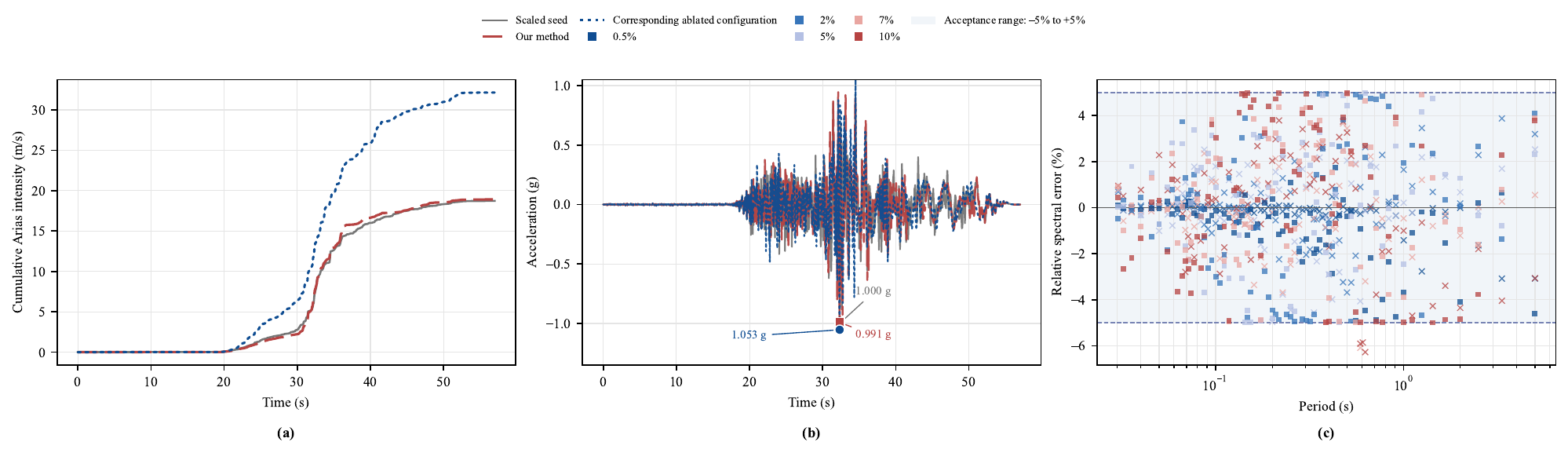}
\caption{Ablation evidence linking each configurable module to its intended observable. (a) Arias modules preserve the cumulative intensity history. (b) PGA-AL enforcement controls the zero-period ordinate; circles identify the PGA values. (c) Extreme-ordinate enforcement confines local spectral errors to the shaded $\pm5\%$ feasibility interval.}
\label{fig:constraintablation}
\end{figure}

Fig.~\ref{fig:constraintablation} and Table~\ref{tab:ablation} test whether each configurable module controls the observable for which it was introduced. The first comparison concerns energy evolution. With the Arias modules active, the adjusted cumulative Arias-intensity history remains close to that of the scaled seed [Fig.~\ref{fig:constraintablation}(a)]. Removing these terms still permits a small spectral error, but the cumulative intensity increases substantially. Accordingly, the total Arias-intensity error rises from 0.98\% to 71.65\% (Table~\ref{tab:ablation}). Spectrum matching alone therefore does not preserve either the energy magnitude or its accumulation pattern when those properties are not explicitly controlled.

A similar effect is observed for PGA in Fig.~\ref{fig:constraintablation}(b). Without the PGA and worst-point spectral constraints, the adjusted motion reaches a PGA of approximately $1.053g$, whereas the complete configuration gives $0.991g$ for the $1.0g$ target. Correspondingly, the PGA error decreases from 5.28\% to 0.87\% when the PGA constraint is activated. This result confirms that minimizing spectral discrepancies at nonzero oscillator periods does not automatically control the zero-period ordinate. PGA therefore requires an explicit constraint when preservation of the target peak acceleration is part of the matching task.

Fig.~\ref{fig:constraintablation}(c) further distinguishes average spectral fidelity from local spectral feasibility. Without the worst-point spectral constraint, the overall MARE remains relatively small at 1.12\%, yet several period--damping ordinates exceed the prescribed $\pm5\%$ error range, resulting in a maximum relative error of 6.28\%. With the extreme-ordinate constraint activated, all controlled ordinates are confined within the prescribed error range and the maximum error is reduced to 5.00\%. Hence, a small average spectral error does not preclude localized violations at individual periods, whereas explicit control of the largest spectral discrepancies complements the distributed spectral objectives by enforcing pointwise feasibility.

Table~\ref{tab:ablation} also shows the optimization trade-off introduced by the additional modules. The complete method gives a MARE of 2.25\%, compared with 0.61\% and 1.12\% for the two reduced configurations. This increase should not be interpreted as degraded spectral-matching performance. Rather, it reflects the expected trade-off of a constrained multi-objective problem: once the optimizer is required to simultaneously control Arias intensity, PGA, and local extreme spectral errors, the feasible search space becomes more restrictive and the available waveform-adjustment freedom can no longer be devoted exclusively to minimizing the average spectral error. The resulting solution therefore represents a compromise among distributed spectral fidelity, pointwise spectral feasibility, and preservation of the selected waveform attributes.

\begin{table}[H]
\caption{Constraint-ablation results for the five-damping benchmark.}
\label{tab:ablation}
\centering
\small
\resizebox{\linewidth}{!}{%
\begin{tabular}{lrrrrr}
\toprule
Configuration & Iterations & $E_{\max}$ (\%) & MARE (\%) & PGA error (\%) & Arias error (\%)\\
\midrule
Complete method & 25,000 & 5.00 & 2.25 & 0.87 & 0.98\\
Without Arias constraints & 25,000 & 4.71 & 0.61 & 0.99 & 71.65\\
Without worst-point AL and PGA-AL & 25,000 & 6.28 & 1.12 & 5.28 & 1.00\\
\bottomrule
\end{tabular}
}
\end{table}

\subsection{Terminal projection and pathwise kinematic control}

Terminal and pathwise control are assessed using three configurations: no terminal or pathwise control, terminal projection alone, and terminal projection combined with the displacement-path constraint.

\begin{figure}[H]
\centering
\includegraphics[width=\linewidth]{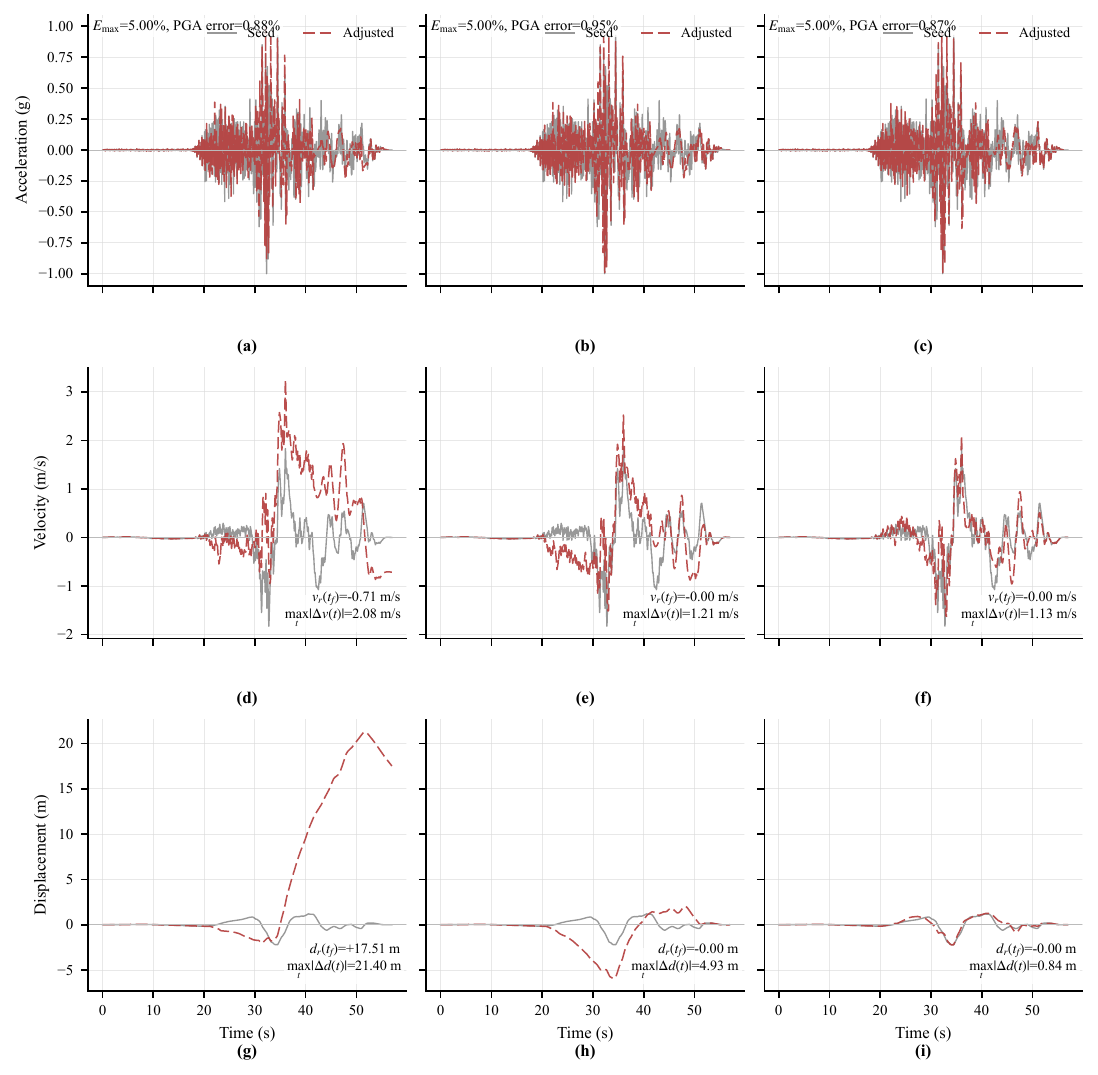}
\caption{Sequential effects of terminal and pathwise kinematic control. (a--c) Acceleration histories without kinematic control, with terminal velocity--displacement projection, and with both terminal projection and the displacement-path constraint, respectively. (d--f) Corresponding velocity histories. (g--i) Corresponding displacement histories. Seed histories are solid and adjusted histories are long dashed; annotations report the spectral and PGA errors together with terminal and peak correction-induced kinematic quantities.}
\label{fig:baseline}
\end{figure}

Fig.~\ref{fig:baseline} isolates the two stages of kinematic control by comparing no control, terminal projection alone, and terminal projection combined with the displacement-path constraint. The columns add these mechanisms sequentially, whereas the rows trace their consequences from acceleration to integrated velocity and displacement. Because the three configurations achieve nearly identical spectral and PGA errors, differences in the lower rows can be attributed primarily to the imposed kinematic controls.

Without any kinematic control, as shown in Figs.~\ref{fig:baseline}(a), (d), and (g), the adjusted motion still achieves a maximum spectral error of approximately 5\%; however, small low-frequency components introduced by the waveform correction accumulate through integration and produce pronounced velocity and displacement drift. The terminal velocity reaches $-0.713$~m/s, the terminal displacement reaches 17.514~m, and the maximum correction-induced displacement reaches 21.397~m. This result demonstrates that satisfactory response-spectrum matching alone does not guarantee acceptable baseline behavior of the integrated motion.

When the hard terminal velocity--displacement projection is activated [Figs.~\ref{fig:baseline}(b), (e), and (h)], each correction is projected onto the feasible subspace with zero correction-induced terminal velocity and displacement increments. Consequently, the terminal velocity and displacement are reduced to $-8.81\times10^{-5}$~m/s and $-5.69\times10^{-4}$~m, respectively, effectively eliminating the terminal drift introduced by spectral adjustment. Terminal compatibility alone, however, constrains only the final integrated state and does not restrict intermediate excursions. Although the final displacement is nearly zero, the maximum correction-induced displacement remains 4.927~m, indicating that a substantial intermediate displacement excursion can still occur before the trajectory returns to its terminal value.

After the displacement-path constraint is additionally activated [Figs.~\ref{fig:baseline}(c), (f), and (i)], the terminal velocity and displacement remain close to zero, while the maximum correction-induced displacement is further reduced from 4.927~m to 0.842~m. The two mechanisms therefore play complementary roles: the hard projection guarantees that the spectral-matching correction does not alter the terminal velocity and displacement of the seed motion, whereas the soft pathwise constraint limits excessive displacement excursions throughout the complete integration history. Their combination provides effective control of adjustment-induced baseline drift while retaining essentially the same spectral-matching accuracy.

\subsection{Integrated five-damping benchmark}

\begin{figure}[H]
\centering
\includegraphics[width=0.82\linewidth]{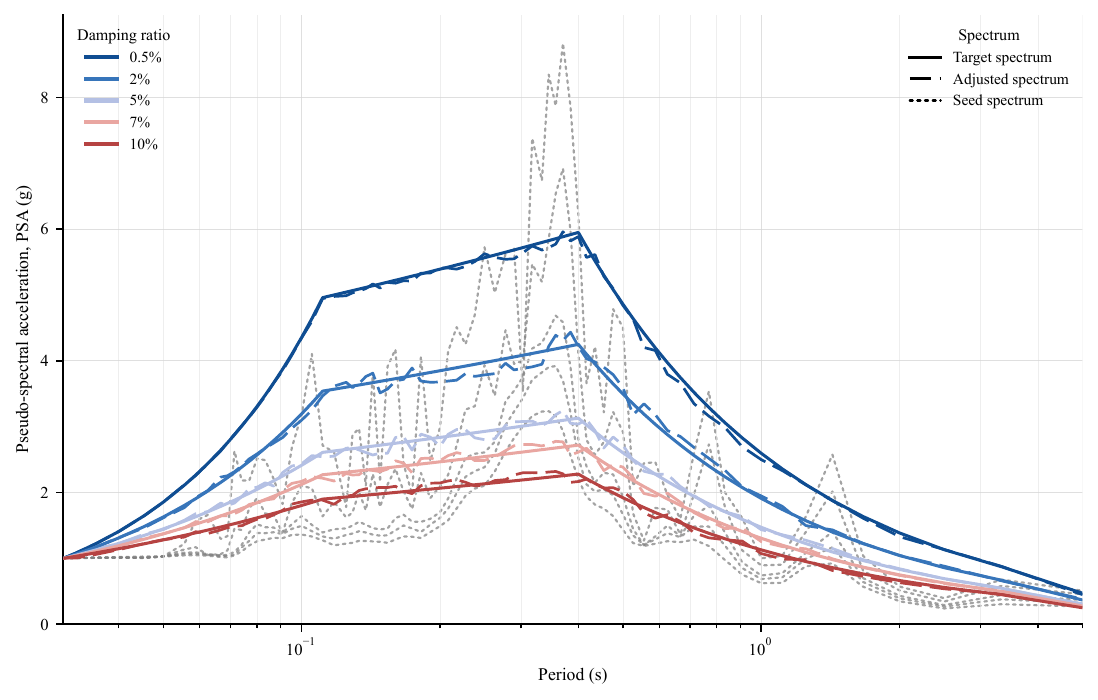}
\caption{Integrated five-damping benchmark after activating the complete module set. One adjusted waveform brings the pseudo-spectral accelerations toward five damping-dependent targets. Target spectra are solid, adjusted spectra are long dashed, and seed spectra are grey short dotted curves.}
\label{fig:strictmulti}
\end{figure}

Having established the role of the individual modules, we next evaluate their integrated behavior on the five-damping target. The seed spectra in Fig.~\ref{fig:strictmulti} deviate from their targets by different amounts across both period and damping ratio. After inverse optimization, one adjusted waveform brings all five spectra toward their respective targets over the full control range. Across the 375 period--damping ordinates, the maximum relative error is 5.00\%, the MARE is 2.25\%, and the RMS error is 2.79\%. Thus, the method coordinates several damping-dependent responses through a single time-domain correction. Read together with the preceding ablations, this result shows that multi-damping agreement is obtained while PGA, Arias intensity, and baseline-related kinematics remain controlled.

\section{Benchmark comparison with established spectrum-matching methods}\label{sec:comparison}

\subsection{Common benchmark and evaluation protocol}

Al Atik and Abrahamson~\cite{alatik2010} introduced an improved adjustment wavelet but did not report a specific seed-record--target-spectrum pair that could be reproduced as a common benchmark. Their procedure is a direct development of the time-domain spectral-matching framework implemented in RspMatch and subsequently advanced by Hancock et al.~\cite{hancock2006}. By contrast, Hancock et al. documented both the input motion and target spectra in a benchmark that has since been reused in comparative studies of spectrum-matching methods~\cite{harati2021}. The present comparison therefore adopts the same benchmark so that differences between the Al Atik implementation and the proposed method can be evaluated from a shared, traceable starting point. The seed is the 1989 $M_w=6.9$ Loma Prieta record at Diamond Heights (PEER NGA record 00794T), recorded approximately 71~km from the fault rupture and scaled by 3.2 in the original example. The target represents a stiff-soil site approximately 10~km from an $M_w=7$ earthquake. The single-damping case uses the 5\% target over 0.05--5.0~s; the multi-damping case uses the 5\%, 10\%, 20\%, and 30\% targets.

Comparison fairness is established at the data and evaluation levels. Each method receives the same scaled seed, target spectra, damping set, and control periods for a given case. The exported waveforms are then evaluated by one independent response-spectrum routine on the same grid, so the entries in Table~\ref{tab:methodcomparison} do not mix method-internal spectrum calculations. The algorithms do not optimize identical secondary objectives: PGA and Arias control are explicit only in the proposed configuration. Spectral errors are therefore directly comparable, whereas differences in PGA, Arias accumulation, velocity, and displacement document the consequences of each method's design rather than an equal-objective competition.

\subsection{Single-damping matching}

\begin{figure}[H]
\centering
\includegraphics[width=\linewidth]{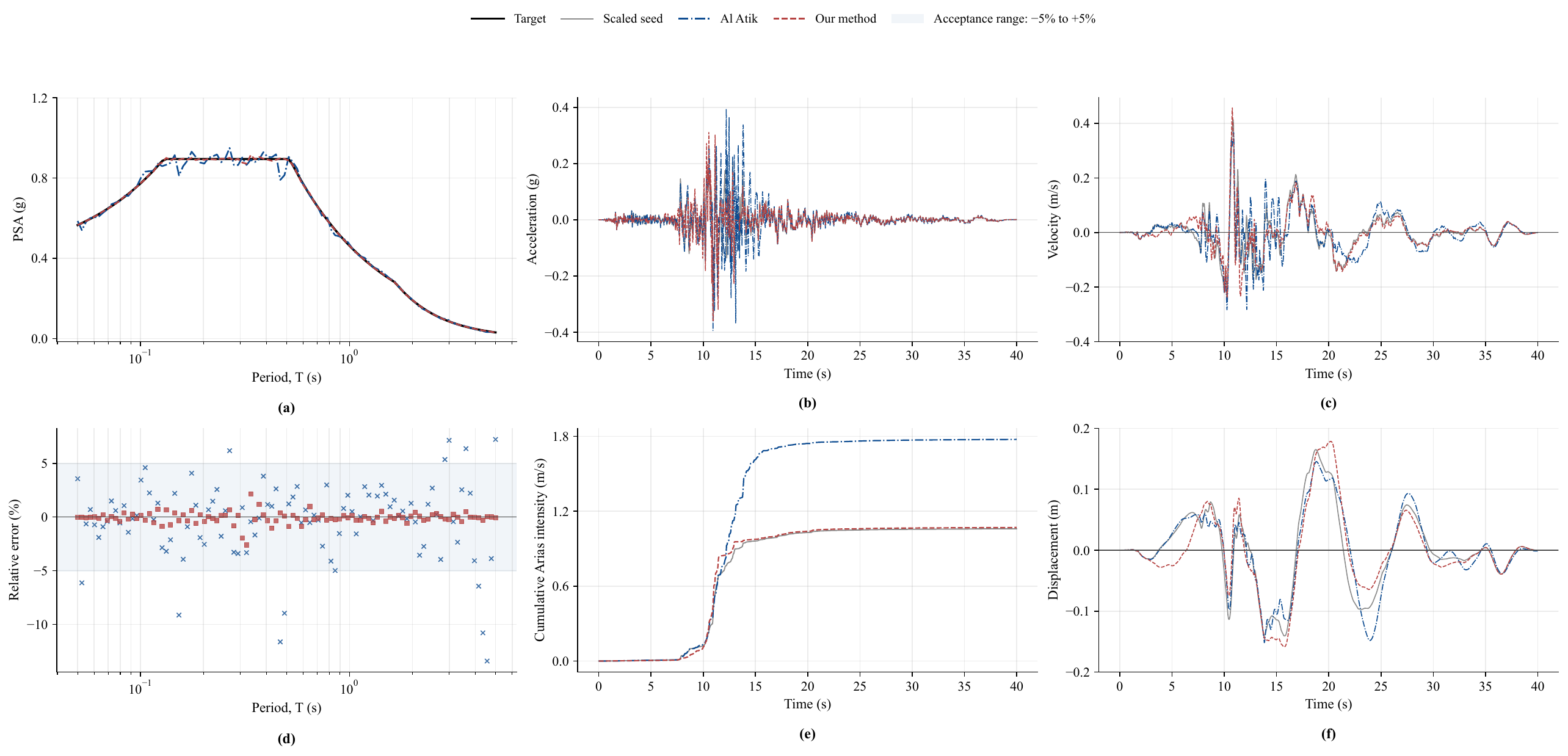}
\caption{Single-damping comparison between the Al Atik and proposed methods on common input data, linking spectral accuracy to time-domain consequences. (a) Response spectra. (b) Acceleration histories. (c) Velocity histories. (d) Pointwise relative spectral errors, with shading denoting the adopted acceptance interval; filled squares denote the proposed method and crosses denote Al Atik. (e) Cumulative Arias-intensity histories. (f) Displacement histories.}
\label{fig:singlecompare}
\end{figure}

The single-damping benchmark examines whether the added waveform constraints compromise the primary spectrum-matching objective. Fig.~\ref{fig:singlecompare} compares the spectral agreement and associated time-domain consequences of the two methods. Both methods start from the same scaled seed and target. The proposed method achieves a MARE of 0.35\%, an RMS error of 0.55\%, and a maximum relative error of 2.61\%; the corresponding Al Atik values are 2.55\%, 3.62\%, and 13.42\%. Consistent with these statistics, the errors in Fig.~\ref{fig:singlecompare}(d) remain closest to zero for the proposed result across the control range.

What is more important in this comparison is that the lower spectral error is not obtained by relaxing the control of other waveform quantities. PGA, Arias intensity, terminal velocity--displacement increments, and the correction-induced displacement path are all explicitly considered in the proposed optimization. The resulting PGA error is 0.96\%, compared with 10.09\% for Al Atik. The cumulative Arias-intensity history in Fig.~\ref{fig:singlecompare}(e) also remains close to that of the scaled seed, while the velocity and displacement histories in Figs.~\ref{fig:singlecompare}(c) and (f) show that the integrated motion is kept under the imposed kinematic controls. Thus, for this single-damping case, the additional constraints do not come at the expense of spectral accuracy; the adjusted motion satisfies a broader set of waveform requirements while yielding smaller spectral errors than the Al Atik result.

\subsection{Multi-damping matching}

\begin{figure}[H]
\centering
\includegraphics[width=\linewidth]{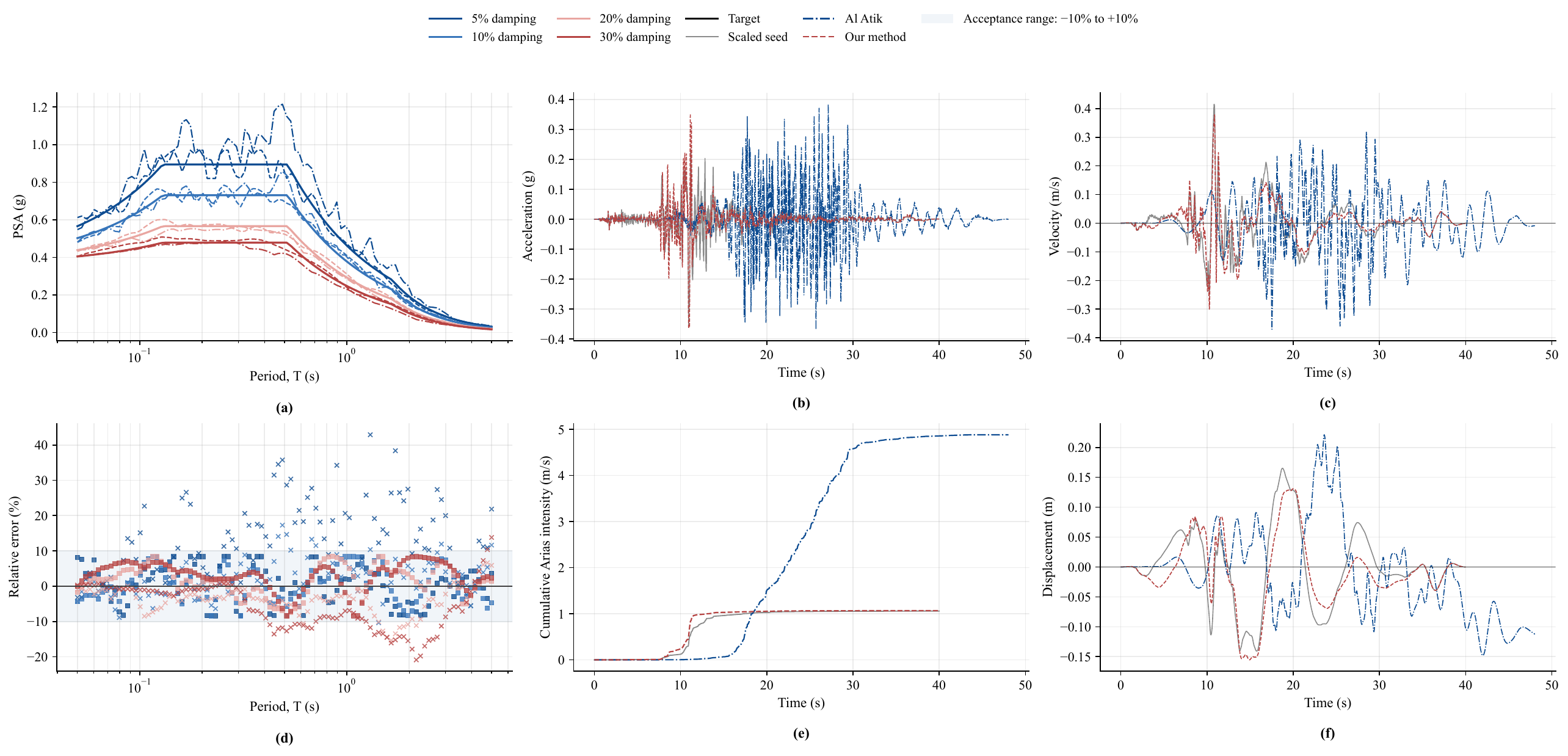}
\caption{Multi-damping comparison between the Al Atik and proposed methods on common input data, linking spectral accuracy to time-domain consequences. (a) Response spectra at four damping ratios. (b) Acceleration histories. (c) Velocity histories. (d) Pointwise relative spectral errors, with the $\pm10\%$ comparison region shaded. (e) Cumulative Arias-intensity histories. (f) Displacement histories.}
\label{fig:multicompare}
\end{figure}

Fig.~\ref{fig:multicompare} repeats the same comparison after the target is expanded from one 5\%-damped spectrum to four spectra at 5\%, 10\%, 20\%, and 30\% damping. A single correction must now coordinate responses across several damping levels, so improvement at one period--damping combination can interfere with another. This stronger coupling increases the errors of both methods relative to the single-damping case.

The increase, however, is markedly different between the methods. The Al Atik result gives a MARE of 7.30\% and a maximum error of 42.91\%, whereas the proposed method limits the corresponding values to 4.19\% and 8.47\%, with an RMS error of 4.91\%. The pointwise distribution in Fig.~\ref{fig:multicompare}(d) shows that the remaining discrepancies are also much more confined across the four damping spectra. This behavior is particularly relevant because the same PGA, Arias-intensity, terminal, and displacement-path controls used in the single-damping case remain active. In other words, the improvement in multi-damping spectral agreement is achieved within a more restrictive optimization problem rather than by allowing larger changes in the other controlled waveform attributes.

This effect is also visible in the time-domain results. The proposed PGA error remains at 1.01\%, compared with 7.33\% for Al Atik. Although the Al Atik result exhibits little terminal baseline drift in the single-damping case, its terminal displacement changes from approximately $-0.0014$~m to $-0.1124$~m in the multi-damping case [Fig.~\ref{fig:multicompare}(f)]. This deterioration occurs because multi-damping matching requires repeated, coupled local corrections, while terminal and pathwise compatibility are not imposed as global constraints on the final Al Atik waveform; small residual low-frequency components can therefore accumulate through double integration. The comparison highlights the main advantage of the proposed formulation: as the spectral target becomes more strongly coupled, multiple waveform and kinematic requirements can still be retained without sacrificing the response-spectrum fit.

\begin{table}[H]
\caption{Common-grid comparison with the Al Atik implementation.}
\label{tab:methodcomparison}
\centering\small
\resizebox{\linewidth}{!}{%
\begin{tabular}{llrrrr}
\toprule
Case & Method & MARE (\%) & RMS (\%) & $E_{\max}$ (\%) & PGA error (\%)\\
\midrule
Single damping & Al Atik & 2.55 & 3.62 & 13.42 & 10.09\\
 & Our method & 0.35 & 0.55 & 2.61 & 0.96\\
\addlinespace
Multiple damping & Al Atik & 7.30 & 10.38 & 42.91 & 7.33\\
 & Our method & 4.19 & 4.91 & 8.47 & 1.01\\
\bottomrule
\end{tabular}}
\end{table}

\section{Engineering applications under configurable acceptance rules}\label{sec:applications}

\subsection{Acceptance-map configuration}

Table~\ref{tab:acceptancemap} orders the two applications by acceptance-map complexity. The first asks for a symmetric pointwise band on one 5\%-damped spectrum. The second combines five common damping spectra, an asymmetric dense-spectrum interval, one-sided and counting rules, and neighborhood checks. In both cases the waveform parameterization, terminal projection, and SDOF operator remain fixed; only the active optimization modules and external checks change. The spectrum, PGA, Arias-intensity, and displacement-path requirements enter gradient-based optimization. In the multi-damping application, differentiable or subdifferentiable violations are also constructed for the under-target count, spectral sum, dense 5\%-damped range, and frequency-neighborhood clauses. Their exact spectral rules are evaluated at optimization checkpoints and participate in the stable-feasibility stopping criterion. Significant duration and component correlation are evaluated only after the three component waveforms have been generated. For both applications, significant duration is
\begin{equation}
 D_{5-75}=t_{75}-t_{5},
\end{equation}
where $t_p$ is the time at which $p$ percent of the final Arias intensity has accumulated. Component independence is evaluated from the zero-lag Pearson correlation coefficients of the acceleration histories, and $\max|\rho|$ is the largest absolute coefficient among H1--H2, H1--V, and H2--V.

\begin{table}[H]
\caption{Acceptance-map configuration for the two engineering applications.}
\label{tab:acceptancemap}
\centering\scriptsize
\resizebox{\linewidth}{!}{%
\begin{tabular}{p{3.3cm}p{4.0cm}p{4.7cm}p{2.5cm}}
\toprule
Requirement & Single-damping application & Multi-damping application & Treatment\\
\midrule
Common-grid response spectrum & $\pm5\%$ at 82 nonzero periods & Five lower envelopes on the common damping grids & Continuous constraint\\
Dense 5\%-damped spectrum & -- & Pointwise range $-10\%$ to $+30\%$ & Continuous constraint\\
PGA & Horizontal $0.07g$; vertical $0.65\times$ horizontal & Horizontal $1.0g$; vertical $2/3g$ & Continuous constraint\\
Under-target count & -- & At most five per common damping spectrum & Differentiable surrogate and exact checkpoint test\\
Spectral sum & -- & Computed sum not below target sum & Continuous constraint and checkpoint test\\
Frequency neighborhood & -- & No all-below-target $\pm10\%$ neighborhood & Differentiable surrogate and exact checkpoint test\\
Significant duration & $D_{5-75}>6$~s & $D_{5-75}>6$~s & External check\\
Component correlation & $\max|\rho|<0.16$ & $\max|\rho|<0.16$ & External check\\
\bottomrule
\end{tabular}
}
\end{table}

\subsection{Single-damping three-component application}

The first application follows the GB/T~50011--2010 (2024 edition) 5\%-damped design spectrum for seismic intensity 8, design basic ground acceleration 0.20$g$, Site Class~II, Design Earthquake Group~1, and the frequent-earthquake level. The horizontal target PGA is 0.07$g$; the vertical target is set to 0.65 times the horizontal target for this case. Following the GB~17741--2025 frequency-spacing specification used here, 82 nonzero periods span 0.02--6.0~s and the PGA point is added.

The active differentiable map consists of the symmetric pointwise spectral constraint, PGA and Arias requirements, terminal projection, and displacement-path control. Significant duration and the three inter-component correlations are not inserted into the loss; they are evaluated as suite-level external checks after the three components have been generated.

\begin{figure}[H]
\centering
\includegraphics[width=\linewidth]{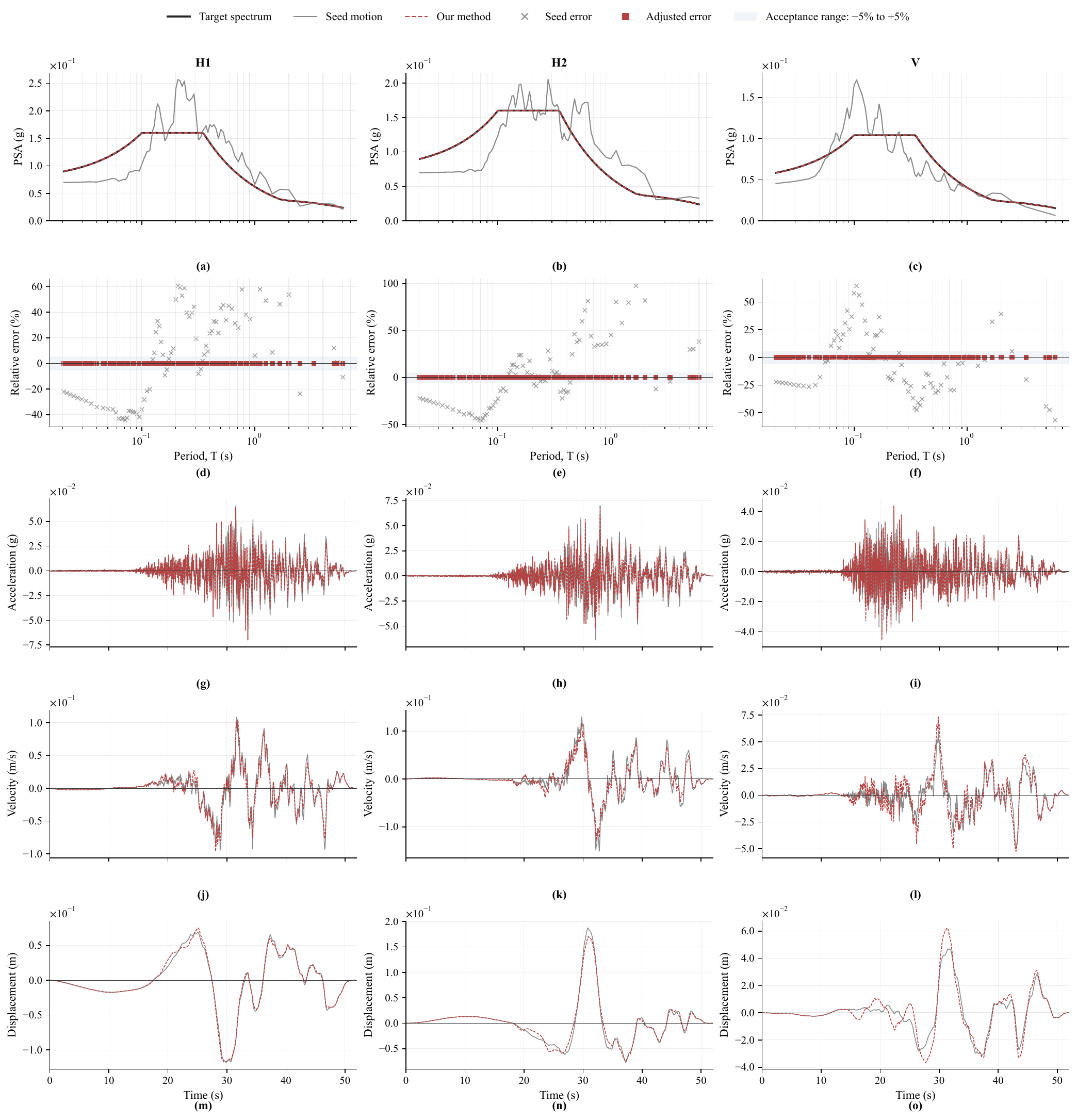}
\caption{Single-damping three-component application using NGA-West2 record RSN~2777. (a--c) Target, scaled-seed, and adjusted 5\%-damped spectra for H1, H2, and V, respectively. (d--f) Corresponding pointwise relative errors within the shaded $-5\%$ to $+5\%$ acceptance interval. (g--i) Acceleration histories. (j--l) Velocity histories. (m--o) Displacement histories.}
\label{fig:gbcase}
\end{figure}

The first application tests whether the configured single-damping rules can be satisfied simultaneously for three components. Fig.~\ref{fig:gbcase} reads from spectral evidence in the upper rows to waveform and kinematic consequences in the lower rows. For the selected NGA-West2 three-component record (RSN~2777), the seed spectra initially contain period-dependent deviations beyond the prescribed $\pm5\%$ tolerance [Figs.~\ref{fig:gbcase}(a)--(f)]. After optimization, the 5\%-damped spectra of H1, H2, and V closely follow their targets, and their pointwise errors approach zero. Table~\ref{tab:gbcase} confirms MARE and RMS errors below 0.01\% for every component and a maximum relative error of 0.01\%.

The spectral agreement is obtained while the PGA- and Arias-intensity-related modules remain active. The PGA errors are 0.98\%, 0.96\%, and 0.58\% for H1, H2, and V, respectively, and the corresponding Arias-intensity errors are 1.00\%, 0.99\%, and 1.00\%. Thus, the nearly exact spectral agreement is not achieved through uncontrolled changes in peak acceleration or total energy. The envelope-capacity multiplier also remains at $\gamma_E=1.0$ for all three components, indicating that the prescribed correction capacity is sufficient for this record set without additional envelope expansion.

Figs.~\ref{fig:gbcase}(g)--(o) further show the acceleration, velocity, and displacement histories before and after adjustment. Most waveform modifications are concentrated within the strong-motion portion of the records, while the overall temporal evolution of the seed motions remains recognizable. The integrated velocity and displacement histories do not exhibit pronounced terminal drift introduced by the spectral adjustment, consistent with the hard terminal velocity--displacement projection. The additional displacement-path constraint limits excessive intermediate excursions during optimization. These constraints do not require the adjusted velocity and displacement histories to reproduce the seed trajectories point by point; accordingly, some changes in PGV and PGD remain in Table~\ref{tab:gbcase} while the terminal and pathwise kinematic requirements are still satisfied.

The significant durations $D_{5-75}$ are 12.93, 13.17, and 15.61~s for H1, H2, and V, respectively, all exceeding the adopted 6-s requirement. Taken together, Fig.~\ref{fig:gbcase} and Table~\ref{tab:gbcase} show that the three-component spectra can be matched to high accuracy while PGA, Arias intensity, and baseline-related kinematic behavior remain simultaneously controlled. This case therefore illustrates how the configurable modules operate jointly under a practical single-damping acceptance map.

\begin{table}[H]
\caption{Results for the single-damping GB/T~50011--2010 (2024 edition)/GB~17741--2025 three-component application. PGV and PGD entries give seed/adjusted values. Percentage errors are reported to two decimal places.}
\label{tab:gbcase}
\centering
\scriptsize
\resizebox{\linewidth}{!}{%
\begin{tabular}{lrrrrrrrr}
\toprule
Comp. & $\gamma_E$ & MARE (\%) & RMS (\%) & $E_{\max}$ (\%) & PGA err. (\%) & Arias err. (\%) & $D_{5-75}$ (s) & PGV (m/s); PGD (m)\\
\midrule
H1 & 1.00 & $<0.01$ & $<0.01$ & 0.01 & 0.98 & 1.00 & 12.93 & 0.1090/0.1178; 0.1180/0.1815\\
H2 & 1.00 & $<0.01$ & $<0.01$ & 0.01 & 0.96 & 0.99 & 13.17 & 0.1510/0.1303; 0.1873/0.2181\\
V  & 1.00 & $<0.01$ & $<0.01$ & 0.01 & 0.58 & 1.00 & 15.61 & 0.0609/0.0660; 0.0469/0.0706\\
\bottomrule
\end{tabular}
}
\end{table}

\subsection{Multi-damping three-component application}

The second application follows the common-spectrum and additional dense 5\%-damping acceptance framework of GB~50267--2019 used in this study~\cite{gb50267}. The common grid contains 75 nonzero points over 0.2--33~Hz at damping ratios of 0.5\%, 2\%, 5\%, 7\%, and 10\%, plus PGA. The additional dense 5\%-damped grid contains 271 nonzero points over 0.1--50~Hz. The time step is 0.01~s, the record length is 62~s, and horizontal and vertical target PGAs are $1.0g$ and $2/3g$, respectively. Here, the below-target count is evaluated separately for each damping ratio on the common grid; the spectral-sum ratio is the sum of computed ordinates divided by the corresponding target sum; the common minimum is the smallest signed relative error on that grid; and the dense range is the minimum-to-maximum signed error on the additional dense 5\%-damped grid. A frequency-neighborhood failure denotes a control point for which the computed spectrum is entirely below the target within the prescribed $\pm10\%$ frequency band.

This configuration uses differentiable lower-envelope, PGA, Arias, terminal, and pathwise modules to guide waveform generation. Under-target counts, spectral sums, dense-range membership, and frequency-neighborhood conditions are represented by continuous or subdifferentiable violations during optimization, and their exact rules participate in checkpoint-level stopping. Significant duration and component correlation remain suite-level external checks after the three components have been generated.

\begin{figure}[H]
\centering
\includegraphics[width=\linewidth]{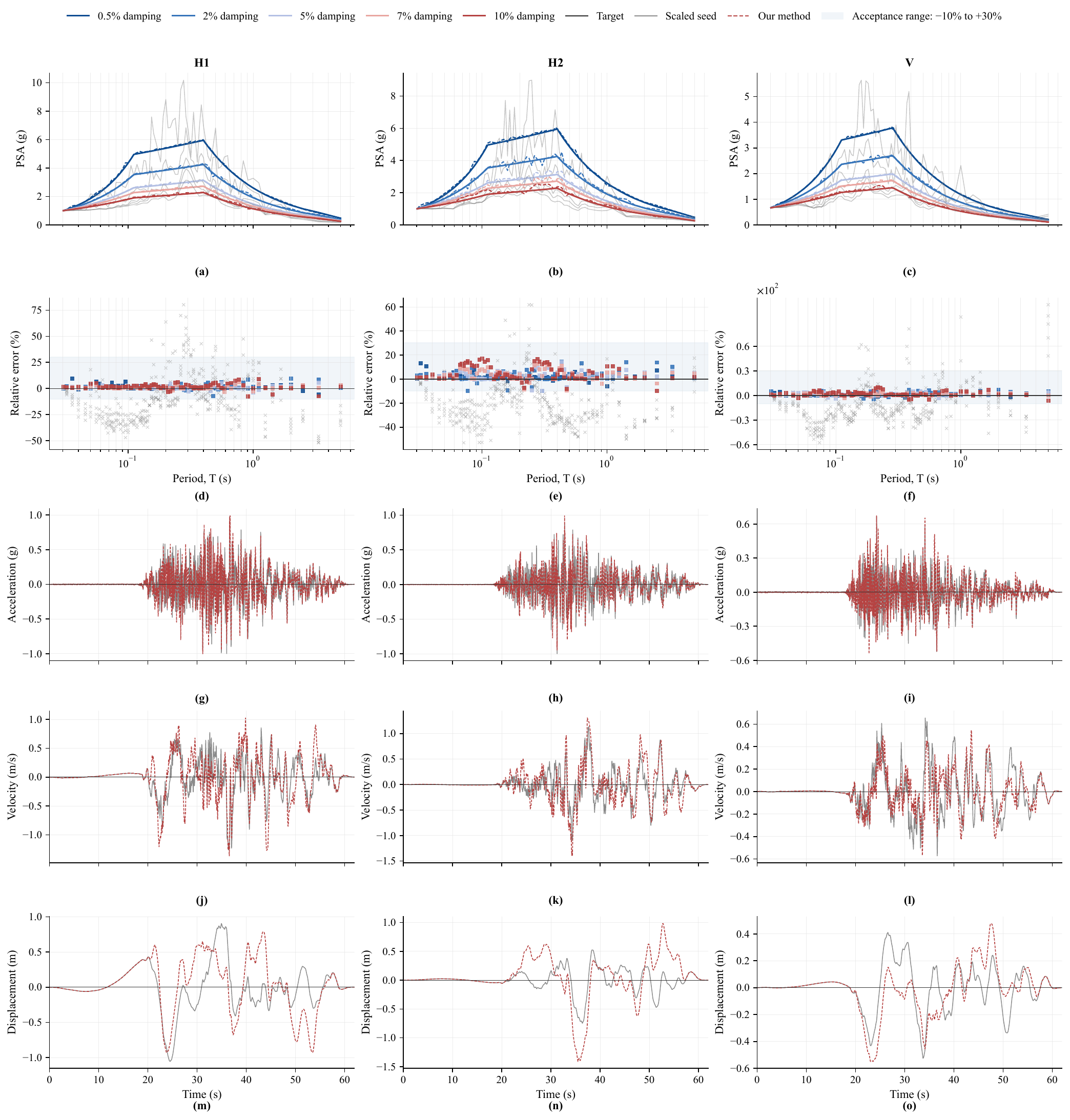}
\caption{Representative multi-damping three-component application. (a--c) Target, scaled-seed, and adjusted spectra for H1, H2, and V, respectively, at five damping ratios. (d--f) Corresponding pointwise relative errors; the shaded $-10\%$ to $+30\%$ band represents only the dense 5\%-damping criterion. (g--i) Acceleration histories. (j--l) Velocity histories. (m--o) Displacement histories.}
\label{fig:sample20}
\end{figure}

The second application asks whether the same fixed computational engine can satisfy a more restrictive multi-damping acceptance map. Fig.~\ref{fig:sample20} again links spectral evidence in the upper rows to time-domain consequences below. For the selected multi-damping three-component input, the H1, H2, and V seed spectra initially show pronounced period- and damping-dependent over- and underestimation. After optimization, one adjusted waveform per component moves all five damping-dependent spectra toward their respective targets [Figs.~\ref{fig:sample20}(a)--(c)]. The corresponding errors become concentrated near the target level [Figs.~\ref{fig:sample20}(d)--(f)], providing the visual evidence that the responses are coordinated across periods and damping ratios.

Acceptance in this case is not determined by a single average spectral-error measure. Table~\ref{tab:sample20} shows that the numbers of below-target ordinates are 5/5/5/5/5 for both H1 and H2 and 4/5/5/5/5 for V across the five damping ratios, so none exceeds the allowable count of five. The minimum relative errors on the common grid are $-7.73\%$, $-9.88\%$, and $-6.35\%$ for H1, H2, and V, respectively, all remaining above the $-10\%$ lower limit. Their minimum spectral-sum ratios are 1.0113, 1.0156, and 1.0081, each exceeding unity. For the additional dense 5\%-damped spectrum, the error ranges are $-9.92\%$ to $+20.21\%$, $-9.88\%$ to $+13.89\%$, and $-7.82\%$ to $+8.13\%$, respectively, and no frequency-neighborhood failure is observed. The adjusted motions therefore satisfy a feasible region defined jointly by several spectral criteria rather than by mean spectral agreement alone.

Figs.~\ref{fig:sample20}(g)--(o) further show the corresponding acceleration, velocity, and displacement histories. Although substantial waveform modification is required to achieve simultaneous five-damping matching, the PGA, Arias-intensity, and baseline-related kinematic controls remain active throughout the optimization. The PGA errors are only 0.20\%, 0.80\%, and 0.90\% for H1, H2, and V, while the corresponding Arias-intensity errors are 0.79\%, 0.96\%, and 0.96\%. The integrated velocity and displacement histories show no uncontrolled terminal drift introduced by the spectral adjustment, reflecting the combined action of the hard terminal projection and the displacement-path constraint. The latter does not force the adjusted displacement history to reproduce the seed trajectory point by point; instead, it limits excessive correction-induced excursions during the matching process.

Compared with the single-damping application, this multi-damping case requires greater correction capacity, with final envelope multipliers $\gamma_E$ of 2.25, 2.50, and 1.75 for H1, H2, and V, respectively. Nevertheless, the corresponding $D_{5-75}$ values remain 16.46, 15.89, and 14.97~s, and the maximum zero-lag inter-component correlation is $0.11118<0.16$. The results of this multi-damping application therefore show that the proposed formulation can accommodate a substantially more restrictive acceptance map---including five damping-dependent spectra, under-target counts, spectral-sum requirements, an additional dense spectrum, and frequency-neighborhood checks---while simultaneously retaining control of PGA, Arias intensity, and baseline-related kinematic behavior.

\begin{table}[H]
\caption{Results for the representative multi-damping three-component application. Below-target counts are listed in the order 0.5\%, 2\%, 5\%, 7\%, and 10\%. The dense range refers to the additional 5\%-damped spectrum.}
\label{tab:sample20}
\centering
\scriptsize
\resizebox{\linewidth}{!}{%
\begin{tabular}{lrrrrrrrrrl}
\toprule
Comp. & $\gamma_E$ & Common min. & Below-target counts & Min. sum ratio & Dense 5\% range & Neighborhood failures & $D_{5-75}$ (s) & PGA/Arias err. & $\max|\rho|$ & Status\\
\midrule
H1 & 2.25 & $-7.73\%$ & 5/5/5/5/5 & 1.0113 & $-9.92$--$20.21\%$ & 0 & 16.46 & 0.20/0.79\% & \multirow{3}{*}{0.11118} & Pass\\
H2 & 2.50 & $-9.88\%$ & 5/5/5/5/5 & 1.0156 & $-9.88$--$13.89\%$ & 0 & 15.89 & 0.80/0.96\% & & Pass\\
V  & 1.75 & $-6.35\%$ & 4/5/5/5/5 & 1.0081 & $-7.82$--$8.13\%$ & 0 & 14.97 & 0.90/0.96\% & & Pass\\
\bottomrule
\end{tabular}
}
\end{table}

\subsection{Cross-case interpretation}

The first application uses a dense but mathematically simple acceptance map: one damping ratio and one symmetric pointwise tolerance. The second is more heterogeneous because five common spectra must satisfy lower-envelope and counting conditions while an additional dense spectrum must remain inside an asymmetric interval. Their results show that the same differentiable engine can support both configurations without changing the oscillator solver or waveform representation.

In the implemented workflow, continuous or subdifferentiable spectral violations supply optimization gradients, while the corresponding exact count, sum, and neighborhood rules are applied at periodic feasibility checkpoints. Duration and component correlation remain post-generation external checks because they are not part of the record-wise optimization graph. Passing these two maps demonstrates feasibility for the investigated records and targets; it does not imply that every code clause or seed record is automatically feasible. A new rule must first be classified by mathematical form and then assigned to a differentiable module, a checkpoint test, or an external verification step.

\section{Discussion}\label{sec:discussion}

\subsection{From prescribed spectral correction to differentiable waveform inverse design}

Conventional spectrum-matching methods generally prescribe the form and placement of correction functions and then update their parameters according to the current spectral discrepancy. The present formulation reorganizes this process by defining the required engineering observables first and obtaining the waveform correction directly from gradients propagated through the SDOF dynamics. This distinction becomes important when spectrum matching is accompanied by additional requirements. The ablation results show that small average spectral errors do not automatically control Arias intensity, PGA, local extreme ordinates, or integrated-motion behavior. The benchmark results further show that these additional constraints can remain active without sacrificing spectral accuracy. The main contribution is therefore not simply a new correction waveform. It is a differentiable inverse-design formulation in which spectral targets and selected waveform attributes are optimized simultaneously within the same candidate motion.

\subsection{Configurability for heterogeneous engineering requirements}

Engineering acceptance requirements differ substantially in mathematical form. Continuous quantities such as pointwise spectral bounds, PGA, Arias intensity, and correction-induced displacement can be incorporated directly into gradient-based optimization. Discrete spectral clauses, including under-target counts, spectral sums, and frequency-neighborhood rules, are represented by differentiable surrogates and verified by exact checkpoint tests. Significant duration and component correlation remain explicit post-generation acceptance checks. The two applications demonstrate how these requirements can be assembled without modifying the waveform parameterization or differentiable SDOF solver. The same principle can be extended to other engineering standards. A new clause is first classified by mathematical form and then represented as an objective, a continuous constraint, a checkpoint test, or an external check. Different code-specific requirements can thus be accommodated through module configuration rather than redesign of the underlying spectrum-matching algorithm.

\subsection{Limitations and future work}

The inverse-design problem is nonconvex, and response-spectrum extrema are only piecewise differentiable: a change in the response-controlling time sample can produce a nonsmooth local gradient. These properties preclude a formal guarantee of global optimality. They did not, however, constitute an evident practical obstacle in the cases investigated here. The optimization is high dimensional and underdetermined: the vector $\bm r\in\mathbb R^N$ contains thousands of time-domain waveform parameters, whereas the complete five-damping target contains 375 spectral ordinates together with a substantially smaller set of additional engineering requirements. The resulting non-uniqueness supplies waveform directions through which spectral, PGA, Arias-intensity, and kinematic requirements can be coordinated. Studies of high-dimensional nonconvex optimization have emphasized that saddle points and flat regions, rather than poor local minima alone, can govern practical optimization behavior \cite{dauphin2014}, while analyses of overparameterized nonlinear systems illustrate how excess parameterization can alter the loss landscape under specific regularity conditions \cite{liu2022loss}. These results are not directly transferable as convergence theorems for the present waveform problem. Nevertheless, the available waveform directions, together with multi-point spectral control, adaptive objective balancing, augmented-Lagrangian updates, and envelope-capacity expansion, provide a plausible explanation for the stable optimization behavior observed here. This interpretation is empirical and structural; it neither establishes global convergence nor guarantees avoidance of unfavorable stationary points.

The present formulation controls only the waveform quantities represented by its active modules. PGA, Arias intensity and its accumulation pattern, terminal velocity--displacement increments, and correction-induced displacement are considered, whereas Fourier amplitude, phase evolution, time--frequency characteristics, pulse-related features, and other waveform descriptors are not directly constrained. Future work should therefore prioritize Fourier-, phase-, time--frequency-, and pulse-aware differentiable modules, together with improved treatment of nonsmooth spectral gradients and more efficient optimization. Nonlinear structural-response quantities can likewise be introduced as additional differentiable objectives or constraints when required.

\section{Conclusions}

This study formulates response-spectrum matching as differentiable waveform inverse design using a fixed SDOF dynamics engine and configurable optimization modules. The differentiable spectrum operator reproduces an exact solution of the equation of motion with a maximum relative difference of 0.002\%. This agreement provides an accurate forward model for gradient-based waveform optimization. Ablation studies further show that spectral fidelity, Arias intensity, PGA, extreme-ordinate feasibility, terminal compatibility, and displacement-path control address distinct aspects of the matching problem.

The complete framework simultaneously matches five damping-dependent spectra over 375 ordinates while retaining the selected peak, energy, and kinematic requirements. On the common benchmark, it achieves maximum spectral errors of 2.61\% and 8.47\% for the single- and multi-damping cases, respectively, while the additional waveform constraints remain active.

The three-component applications further show that substantially different engineering acceptance requirements can be handled by reconfiguring the active objectives, constraints, checkpoint tests, and external checks. The underlying differentiable dynamics remain unchanged. The high-dimensional waveform parameterization provides substantial freedom for accommodating multiple response and waveform constraints within the same inverse-design problem. This modular organization provides a general route for incorporating additional engineering requirements into response-spectrum matching. Within the demonstrated scope, it can be extended to Fourier-, phase-, time--frequency-, or structural-response-based constraints.

\section*{Declaration of competing interest}
The authors declare that they have no known competing financial interests or personal relationships that could have appeared to influence the work reported in this paper.

\section*{Data availability}
The data used in this study are available from the NGA-West2 database at \url{https://peer.berkeley.edu/ngawest2}. The source code will be made publicly available at \url{https://github.com/hu567/DiffSpecMatch} after publication.

\section*{Acknowledgements}
The authors gratefully acknowledge financial support from the National Natural Science Foundation of China (Nos. U2539204 and 52192675).

\bibliographystyle{elsarticle-num}
\bibliography{references}

\end{document}